\pdfoutput=1
\documentclass[runningheads]{llncs}

\usepackage{mdframed}

\usepackage{amsmath,amsfonts, amssymb}

\usepackage{graphicx}
\graphicspath{{./figures/}}
\usepackage{wrapfig}
\usepackage{nameref}
\usepackage{stmaryrd}
\usepackage[usenames,dvipsnames]{xcolor}
\usepackage{xfrac}
\usepackage{xspace}
\usepackage[inline]{enumitem}
\usepackage{cancel}
\usepackage{csquotes}
\usepackage{adjustbox}
\usepackage{subcaption}
\usepackage{nicefrac}
\usepackage{xfrac}
\usepackage{mathtools}
\usepackage{booktabs}
\usepackage{multirow}
\usepackage{listings}
\usepackage{marvosym}
\usepackage{fontawesome}
\usepackage{pifont}
\usepackage{microtype}
\usepackage{xifthen}
\usepackage{makecell}
\usepackage{array}

\usepackage[ruled,vlined,linesnumbered]{algorithm2e}
\SetKwRepeat{Do}{do}{while}

\SetCommentSty{mycommfont}

\usepackage{tensor}
\usepackage{tikz}
\usetikzlibrary{patterns,arrows,backgrounds,calc,shapes,shadows,decorations.pathmorphing,decorations.pathreplacing,automata,shapes.multipart,positioning,shapes.geometric,fit,circuits,trees,shapes.gates.logic.US,fit,decorations.markings}

\usepackage[font=small,labelfont=bf]{caption}
\usepackage[textwidth=4cm]{todonotes}
\usepackage{pgfplots}
\pgfplotsset{compat=newest}

\makeatletter
\RequirePackage[bookmarks,unicode,colorlinks=true]{hyperref}%
\def\@citecolor{blue}%
\def\@urlcolor{blue}%
\def\@linkcolor{RedViolet}%

\def\orcidID#1{\smash{\href{http://orcid.org/#1}{\protect\raisebox{-1.25pt}{\protect\includegraphics{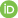}}}}}
\makeatother

\def\ackname{Data-Availability Statement}%
\def\acknowledgement{\par\addvspace{17pt}\small\rmfamily
	\trivlist\if!\ackname!\item[]\else
	\item[\hskip\labelsep
	{\bfseries\ackname}]\fi}

\usepackage[capitalize]{cleveref}
\crefname{lemma}{Lem.}{Lem.}
\crefname{example}{Ex.}{Ex.}
\crefname{equation}{Eq.}{Eq.}
\crefname{section}{Sect.}{Sect.}
\crefname{appendix}{Appx.}{Appx.}
\crefname{definition}{Def.}{Def.}
\crefname{theorem}{Thm.}{Thm.}
\crefname{proposition}{Prop.}{Prop.}
\crefname{corollary}{Cor.}{Cor.}
\crefname{algorithm}{Alg.}{Alg.}
\crefname{figure}{Fig.}{Fig.}
\crefname{table}{Tab.}{Tab.}
\usepackage{placeins}

\definecolor{color1}{RGB}{55,126,184} 
\definecolor{color2}{RGB}{228,26,28} 
\definecolor{color3}{RGB}{77,175,74} 
\definecolor{color4}{RGB}{152,78,163} 
\definecolor{color5}{RGB}{255,127,0} 
\definecolor{color6}{rgb}{0.5, 1.0, 0.83} 
\definecolor{color7}{rgb}{1.0, 0.0, 1.0} 
\definecolor{color8}{rgb}{0.66, 0.66, 0.66} 

\newlength{\scatterplotsize}
\newcommand{\scatterplotstorm}[6]{%
	\begin{tikzpicture}
	\begin{axis}[
	width=\scatterplotsize,
	height=\scatterplotsize,
	axis equal image,
	xmin=0.04,
	ymin=0.04,
	ymax=22000,
	xmax=22000,
	xmode=log,
	ymode=log,
	axis x line=bottom,
	axis y line=left,
	xtick={1,5,20,100,1000,3000},
	xticklabels={1,5,20,100,1000,3000},
	extra x ticks = {10000},
	extra x tick labels = {MO/TO},
	extra x tick style = {grid = major},
	ytick={1,5,20,100,1000,3000},
	yticklabels={1,5,20,100,1000,3000},
	extra y ticks = {10000},
	extra y tick labels = {MO/TO},
	extra y tick style = {grid = major},
	xlabel={#3},
	xlabel style={font=\small,yshift=18pt},
	ylabel={#5},
	ylabel style={font=\small,yshift=-0.55cm},
	yticklabel style={font=\tiny},
	xticklabel style={rotate=290,anchor=west,font=\tiny},
	legend pos=north west,
	legend columns=-1,
	legend style={nodes={scale=0.75, transform shape},inner sep=1.5pt,yshift=0.67cm,xshift=0.8cm},
	]
	
	\addplot[
	scatter,
	only marks,
	scatter/classes={
		dtmc={mark=*,color1,mark size=1}
	},
	scatter src=explicit symbolic
	]%
	table [col sep=semicolon,x=#2,y=#4,meta=Type] {#1};
	\ifthenelse{\NOT\equal{#6}{false}}{\legend{DTMC, CTMC, MDP, MA, PTA}}{}
	\addplot[no marks] coordinates {(0.01,0.01) (10000,10000) };
	\addplot[no marks, densely dotted] coordinates {(0.01,0.1) (1000,10000)};
	\addplot[no marks, densely dotted] coordinates {(0.1,0.01) (10000,1000)};
	\end{axis}
	\end{tikzpicture}
}

\newcommand{\scatterplotabsynth}[6]{%
	\begin{tikzpicture}
		\begin{axis}[
			width=\scatterplotsize,
			height=\scatterplotsize,
			axis equal image,
			xmin=0.001,
			ymin=0.001,
			ymax=950,
			xmax=950,
			xmode=log,
			ymode=log,
			axis x line=bottom,
			axis y line=left,
			xtick={0.01,0.5,5,50},
			xticklabels={0.01,0.5,5,50},
			extra x ticks = {500},
			extra x tick labels = {\phantom{MO/TO}},
			extra x tick style = {grid = major},
			ytick={0.01,0.5,5,50},
			yticklabels={0.01,0.5,5,50},
			extra y ticks = {500},
			extra y tick labels = {\phantom{MO/TO}},
			extra y tick style = {grid = major},
			xlabel={#3},
			xlabel style={font=\small,yshift=18pt},
			ylabel={#5},
			ylabel style={font=\small,yshift=-0.58cm},
			yticklabel style={font=\tiny},
			xticklabel style={rotate=290,anchor=west,font=\tiny},
			legend pos=north west,
			legend columns=-1,
			legend style={nodes={scale=0.75, transform shape},inner sep=1.5pt,yshift=0.67cm,xshift=0.8cm},
			]
			
			\addplot[
			scatter,
			only marks,
			scatter/classes={
				dtmc={mark=triangle*,color5,mark size=1}
			},
			scatter src=explicit symbolic
			]%
			table [col sep=semicolon,x=#2,y=#4,meta=Type] {#1};
			\ifthenelse{\NOT\equal{#6}{false}}{\legend{DTMC, CTMC, MDP, MA, PTA}}{}
			\addplot[no marks] coordinates {(0.001,0.001) (500,500) };
			\addplot[no marks, densely dotted] coordinates {(0.001,0.01) (50,500)};
			\addplot[no marks, densely dotted] coordinates {(0.01,0.001) (500,50)};
		\end{axis}
	\end{tikzpicture}
}

\newcommand{\scatterplotexist}[6]{%
	\begin{tikzpicture}
	\begin{axis}[
	width=\scatterplotsize,
	height=\scatterplotsize,
	axis equal image,
	xmin=0.04,
	ymin=0.04,
	ymax=900,
	xmax=900,
	xmode=log,
	ymode=log,
	axis x line=bottom,
	axis y line=left,
	xtick={1,5,10,20,40,100,200},
	xticklabels={1,5,10,20,40,100,200},
	extra x ticks = {500},
	extra x tick labels = {MO/TO},
	extra x tick style = {grid = major},
	ytick={1,5,10,20,40,100,200},
	yticklabels={1,5,10,20,40,100,200},
	extra y ticks = {500},
	extra y tick labels = {MO/TO},
	extra y tick style = {grid = major},
	xlabel={#3},
	xlabel style={font=\small,yshift=18pt},
	ylabel={#5},
	ylabel style={font=\small,yshift=-0.58cm},
	yticklabel style={font=\tiny},
	xticklabel style={rotate=290,anchor=west,font=\tiny},
	legend pos=south east,
	legend columns=1,
	legend style={nodes={scale=0.5, transform shape},inner sep=1pt,yshift=0.06cm,xshift=-0.2cm},
	legend cell align={left}
	]
	
	\addplot[
	scatter,
	only marks,
	scatter/classes={
		unsound={mark=diamond*,color3,mark size=1},
		sound={mark=diamond*,color1,mark size=1}
	},
	scatter src=explicit symbolic
	]%
	table [col sep=semicolon,x=#2,y=#4,meta=Type] {#1};
	\ifthenelse{\NOT\equal{#6}{false}}{\legend{sub-inv.\ only\!, sound bounds\!}}{}
	\addplot[no marks] coordinates {(0.01,0.01) (500,500) };
	\addplot[no marks, densely dotted] coordinates {(0.01,0.1) (50,500)};
	\addplot[no marks, densely dotted] coordinates {(0.001,0.1) (5,500)};
	\addplot[no marks, densely dotted] coordinates {(0.1,0.01) (500,50)};
	\end{axis}
	\end{tikzpicture}
}

\newcommand{\scatterplotexistsoundunsound}[6]{%
	\begin{tikzpicture}
		\begin{axis}[
			width=\scatterplotsize,
			height=\scatterplotsize,
			axis equal image,
			xmin=0.04,
			ymin=0.04,
			ymax=900,
			xmax=900,
			xmode=log,
			ymode=log,
			axis x line=bottom,
			axis y line=left,
			xtick={1,5,10,20,40,100,200},
			xticklabels={1,5,10,20,40,100,200},
			extra x ticks = {500},
			extra x tick labels = {MO/TO},
			extra x tick style = {grid = major},
			ytick={1,5,10,20,40,100,200},
			yticklabels={1,5,10,20,40,100,200},
			extra y ticks = {500},
			extra y tick labels = {MO/TO},
			extra y tick style = {grid = major},
			xlabel={#3},
			xlabel style={font=\small,yshift=18pt},
			ylabel={#5},
			ylabel style={font=\small,yshift=-0.58cm},
			yticklabel style={font=\tiny},
			xticklabel style={rotate=290,anchor=west,font=\tiny},
			legend pos=south east,
			legend columns=1,
			legend style={nodes={scale=0.5, transform shape},inner sep=1pt,yshift=0.06cm,xshift=-0.2cm},
			legend cell align={left}
			]
			
			\addplot[
			scatter,
			only marks,
			scatter/classes={
				sound={mark=diamond*,color4,mark size=1}
			},
			scatter src=explicit symbolic
			]%
			table [col sep=semicolon,x=#2,y=#4,meta=Type] {#1};
			\ifthenelse{\NOT\equal{#6}{false}}{\legend{sub-inv.\ only\!, sound bounds\!}}{}
			\addplot[no marks] coordinates {(0.01,0.01) (500,500) };
			\addplot[no marks, densely dotted] coordinates {(0.01,0.1) (50,500)};
			\addplot[no marks, densely dotted] coordinates {(0.1,0.01) (500,50)};
		\end{axis}
	\end{tikzpicture}
}

\newcommand{\toolname}[1]{\textsc{#1}\xspace}
\newcommand{\tool}{\toolname{cegispro2}}
\newcommand{\headername}[1]{\textsc{#1}\xspace}
\newcommand{\expname}[1]{\textsf{#1}\xspace}

\DeclarePairedDelimiter\abs{\lvert}{\rvert}

\newcommand{\generator}[1]{\langle #1 \rangle}

\newcommand{\temp}{T}

\newcommand{\indinvshort}{\mathsf{AdmInv}}

\newcommand{\tvars}{\mathsf{TVars}}
\newcommand{\tval}{\mathfrak{I}}
\newcommand{\tvalapp}[2]{#1\left[#2\right]}
\newcommand{\tvalapps}[1]{\tvalapp{#1}{\tval}}

\newcommand{\satset}{\mathsf{Sat}}
\newcommand{\counterex}[1]{\mathsf{CounterEx}_#1}

\newcommand{\Temp}{\mathsf{TExp}}
\newcommand{\Ntemp}{\mathsf{TnExp}}
\newcommand{\Tb}{B}
\newcommand{\Te}{E}

\newcommand{\verify}{\mathsf{Verify}}
\newcommand{\coopverify}{\mathsf{CVerify}}
\newcommand{\tsynt}{\mathsf{Synt}}

\newcommand{\synthcex}{\ensuremath{S'}}

\newcommand{\rr}{\ensuremath{r}\xspace}

\newcommand{\inv}{\ensuremath{I} \xspace}

\newcommand{\sizeof}[1]{\mid #1 \mid}

\makeatletter
\newcommand{\monus}{\mathbin{\text{\@dotminus}}}
\newcommand{\@dotminus}{%
	\ooalign{\hidewidth\raise1ex\hbox{.}\hidewidth\cr$\m@th-$\cr}%
}
\makeatother

\newcommand{\cmark}{{\ding{51}}}

\newcommand{\sfsymbol}[1]{\textsf{\upshape {#1}}}

\newcommand{\ttsymbol}[1]{\texttt{\upshape {#1}}}

\newcommand{\wpsymbol}{\sfsymbol{wp}}

\renewcommand{\wp}[2]{\wpsymbol\llbracket #1\rrbracket\left(#2\right)}

\newcommand{\cc}{\ensuremath{C}} 
\newcommand{\guard}{\varphi} 
\newcommand{\ee}{\ensuremath{e}} 

\newcommand{\pp}{\ensuremath{p}} 

\newcommand{\Guard}{\guard}
\newcommand{\Body}{\cc}

\newcommand{\SKIP}{\ttsymbol{skip}}

\newcommand{\AssignSymbol}{\mathrel{\textnormal{$\mathtt{\coloneqq}$}}}
\newcommand{\ASSIGN}[2]{\ensuremath{#1 \AssignSymbol #2}}

\newcommand{\AVAILLOC}[1]{\PosNats}
\usepackage{actuarialangle}

\newcommand{\COMPOSE}[2]{\ensuremath{{#1}{\,;}~ {#2}}}
\newcommand{\PCHOICE}[3]{\ensuremath{\left\{\, {#1} \,\right\}\mathrel{\left[\,#2\,\right]}\left\{\, {#3} \,\right\}}}

\newcommand{\IFSYMBOL}{\ensuremath{\textnormal{\texttt{if}}}}

\newcommand{\ELSESYMBOL}{\ensuremath{\textnormal{\texttt{else}}}}

\newcommand{\ITE}[3]{\ensuremath{\IFSYMBOL\,\left(\, {#1} \,\right)\,\left\{\, {#2} \,\right\}\,\ELSESYMBOL\,\left\{\, {#3} \,\right\}}}
\newcommand{\ITNE}[2]{\ensuremath{\IFSYMBOL\,\left(\, {#1} \,\right)\,\left\{\, {#2} \,\right\}}}

\newcommand{\WHILESYMBOL}{\ensuremath{\textnormal{\texttt{while}}}}
\newcommand{\WHILE}[1]{\ensuremath{\WHILESYMBOL \left(\, {#1} \,\right)\left\{\right.}}

\newcommand{\WHILEDO}[2]{\ensuremath{\WHILESYMBOL \left(\, {#1} \,\right)\left\{\, {#2} \,\right\}}}

\newcommand{\varfont}[1]{\textit{#1}}
\newcommand{\varsent}{\ensuremath{\varfont{sent}}\xspace}
\newcommand{\varfailed}{\ensuremath{\varfont{fail}}\xspace}

\newcommand{\Vars}{\ensuremath{\mathsf{Vars}}\xspace}   

\newcommand{\Nats}{\ensuremath{\mathbb{N}}\xspace}

\newcommand{\PosNats}{\ensuremath{\mathbb{N}_{>0}}\xspace}
\newcommand{\Ints}{\ensuremath{\mathbb{Z}}\xspace}
\newcommand{\Rats}{\ensuremath{\mathbb{Q}}\xspace}
\newcommand{\Reals}{\mathbb{R}}
\newcommand{\PosReals}{\mathbb{R}_{\geq 0}}

\newcommand{\PosRealsInf}{\mathbb{R}_{\geq 0}^\infty}

\newcommand{\E}{\mathbb{E}}

\newcommand{\Elin}{\ensuremath{\mathsf{LinExp}}\xspace}

\newcommand{\iverson}[1]{\left[ {#1} \right]}

\newcommand{\subst}[2]{\left[ {#1} \middle/ {#2}\right]}

\newcommand{\charfun}[1]{\Phi_{#1}}

\newcommand{\tcharfun}[1]{\Psi_{#1}}
\newcommand{\ertcharfun}{\Theta}

\newcommand{\lfp}{\sfsymbol{lfp}~}

\newcommand{\eval}[1]{\ensuremath{\llbracket {#1} \rrbracket}}

\newcommand{\States}{S}
\newcommand{\state}{s}
\newcommand{\stateB}{t}
\newcommand{\initialState}{\state_0}

\newcommand{\true}{\mathsf{true}}
\newcommand{\false}{\mathsf{false}}
\newcommand{\mydot}{\text{{\Large\textbf{.}}~}}

\newcommand{\cf}{\text{cf.}\xspace}
\newcommand{\ie}{\text{i.e.}\xspace}
\newcommand{\eg}{\text{e.g.}\xspace}

\newcommand{\qiff}{~\quad\textnormal{iff}\quad~}

\newcommand{\qand}{~\quad\textnormal{and}\quad~}

\newcommand{\qimplies}{~\quad\textnormal{implies}\quad~}
\newcommand{\qqimplies}{\qquad\textnormal{implies}\qquad}

\newcommand{\ppreceq}{~{}\preceq{}~}

\newcommand{\eeq}{~{}={}~}

\newcommand{\pplus}{~{}+{}~}

\newcommand{\mmid}{~{}\mid{}~}
\newcommand{\qmid}{\quad{}\mid{}\quad}

\newcommand{\nneq}{~{}\neq{}~}

\newcommand{\lleq}{~{}\leq{}~}

\newcommand{\iin}{~{}\in{}~}

\newcommand{\wwedge}{~{}\wedge{}~}

\newcommand{\setcomp}[2]{\left\{\, {#1} ~\middle|~ {#2} \,\right\}}

\newcommand{\gray}[1]{\textcolor{gray}{#1}}

\makeatletter
\newcommand{\labeltarget}[1]{\Hy@raisedlink{\hypertarget{#1}{}}}
\makeatother

\def\triangleforqed{\hbox{$\lhd$}}
\makeatletter
\DeclareRobustCommand{\qedTT}{%
	\ifmmode
	\eqno \def\@badmath{$$}
	\let\eqno\relax \let\leqno\relax \let\veqno\relax
	\hbox{\triangleforqed}%
	\else
	\leavevmode\unskip\penalty9999 \hbox{}\nobreak\hfill
	\quad\hbox{\triangleforqed}%
	\fi
}
\makeatother

\newif\ifblindreview

\blindreviewfalse

\begin{document}
	
\title{
	Probabilistic Program Verification via\\Inductive Synthesis of Inductive Invariants%
	\ifblindreview\else\thanks{\setlength{\leftskip}{0em}%
		This research was funded by the ERC AdG FRAPPANT under grant No.\ 787914.}
	\fi
}

\titlerunning{Probabilistic Program Verification via\\Inductive Synthesis
}
%

\authorrunning{%
	\ifblindreview%
	\else%
	K.\ Batz et al.
	\fi
}

\ifblindreview%
\author{}
\institute{}
\else%
\author{Kevin Batz\inst{1}$^{\text{(\Letter)}}$
	\and
	Mingshuai Chen\inst{2}$^{\text{(\Letter)}}$
	\and
	Sebastian~Junges\inst{3}$^{\text{(\Letter)}}$
	\and
	Benjamin Lucien Kaminski\inst{4}$^{\text{(\Letter)}}$
	\and
	Joost-Pieter~Katoen\inst{1}$^{\text{(\Letter)}}$
	\and
	Christoph~Matheja\inst{5}$^{\text{(\Letter)}}$
}

\institute{RWTH Aachen University, Aachen, Germany\\
	\email{\{kevin.batz,katoen\}@cs.rwth-aachen.de}
	\and
	Zhejiang University, Hangzhou, China\\
	\email{m.chen@zju.edu.cn}
	\and
	Radboud University, Nijmegen,  Netherlands\\
	\email{sebastian.junges@ru.nl}
	\and
	Saarland University, Saarbrücken, Germany\\and University College London, London, United Kingdom\\
	\email{kaminski@cs.uni-saarland.de}
	\and
	Technical University of Denmark, Kgs. Lyngby, Denmark\\
	\email{chmat@dtu.dk}
}
\fi
	
\maketitle              
\setlength{\floatsep}{1\baselineskip}
\setlength{\textfloatsep}{1\baselineskip}
\setlength{\intextsep}{1\baselineskip}
\begin{abstract}



Essential tasks for the verification of probabilistic programs include bounding expected outcomes and proving termination in finite expected runtime. We contribute a simple yet effective \emph{inductive synthesis} approach for proving such \emph{quantitative reachability properties} by generating \emph{inductive invariants} on \emph{source-code level}. Our implementation shows promise: It finds invariants for (in)finite-state programs, can beat state-of-the-art probabilistic model checkers, and is competitive with modern tools dedicated to invariant synthesis and expected runtime reasoning.

\end{abstract}
	\section{Introduction}

Reasoning about reachability probabilities is a foundational task in the analysis of randomized systems.
%
Such systems are (possibly infinite-state) \emph{Markov chains}, which are typically described as \emph{probabilistic programs} -- imperative programs that may sample from probability distributions.
%
%
We contribute a method for proving bounds on \emph{quantitative properties} of probabilistic programs, which finds \emph{inductive invariants} on \emph{source-code level} by \emph{inductive synthesis}. 
%
We discuss each of these ingredients below, present our approach with a running example in \Cref{sec:overview}, and defer a detailed discussion of related work to \Cref{sec:related}. 

\smallskip\noindent
1) \emph{Quantitative Reachability Properties.}
We aim to verify properties such as \emph{\enquote{is the probability of reaching an error at most 1\%?}}
More generally, our technique proves bounds on the expected value of a probabilistic program terminating in designated states (see \Cref{sec:overview:reach}).
Various verification problems are ultimately solved by bounding quantitative reachability properties
(\cf~\cite{Baier2008Principles,McIverM05}).
Further examples of such problems include 
\emph{\enquote{does a program terminate with finite expected runtime?}} 
and \emph{\enquote{is the expected sum of program variables x and y at least one?}}

\smallskip\noindent
2) \emph{Inductive Invariants.} 
An inductive invariant is a \emph{certificate} that witnesses a certain quantitative reachability property.
Quantitative (and qualitative) reachability are typically captured as least fixed points 
(\cf \cite{Puterman1994Markov,McIverM05,Baier2008Principles}).
 For upper bounds, this characterization makes it natural to search for a prefixed point -- the inductive invariant -- that, by standard fixed point theory~\cite{tarski1955lattice}, is greater than or equal to the least fixed point.
Our invariants assign every state a quantity. 
If the initial state is assigned a quantity below the desired threshold, then the invariant certifies that the property in question holds. 
We detail quantitative inductive invariants in \Cref{sec:overview:indinvariants}; we adapt our method to lower bound reasoning in \Cref{sec:past_lower}.


\smallskip\noindent
3) \emph{Source-Code Level.} 
We consider probabilistic programs over (potentially unbounded) integer 
variables that conceptually extend while-programs with 
coin flips, see  \eg \Cref{fig:brp}.\footnote{\toolname{Prism} programs can be interpreted as an implicit \texttt{while(not error-state) $\{ \hdots \}$} program -- see \cite{DBLP:conf/cav/HoltzenJVMSB20} for an explicit translation.}
We exploit the program structure to reason about infinite-state (and large finite-state) programs:
We \emph{never} construct a Markov chain but find \emph{symbolic} inductive invariants (mapping from program states to nonnegative reals) on \emph{source-code} level. 
We particularly discover inductive invariants that are piecewise linear, as they can often be verified efficiently.

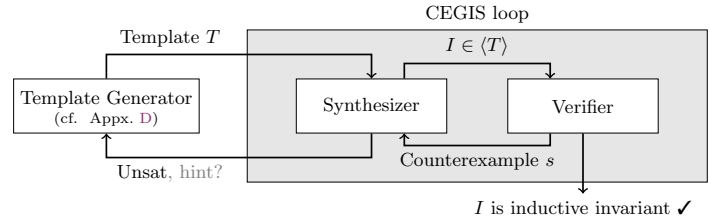
\begin{figure}[t]
\centering
\scalebox{0.8}{
\begin{tikzpicture}
	\node[draw, minimum width=2.5cm, fill=white, minimum height=0.9cm] (synthesiser) {Synthesizer};
	\node[draw, align=center, right=of synthesiser, minimum width=2.5cm, minimum height=0.9cm, fill=white] (verifier) {Verifier};
	\begin{scope}[on background layer]
	\node[draw, fit=(verifier)(synthesiser), inner sep=23pt, label=north:CEGIS loop, fill=black!10] (blackbox) {};
	\end{scope}
	
	\node[below=of verifier] (done) {$I$ is inductive invariant \cmark};
	\node[draw, left=1.6cm of synthesiser, minimum width=2.5cm, minimum height=0.9cm, align=center] (tcreator) {\makecell{Template Generator\\[-.6mm]{\scriptsize (\cf~~\cref{sec:refinement})}\\[-1mm]}};
	
	\draw[->,thick] (verifier) -- node[] {} (done);

	\draw[->,thick] (verifier.220) -- +(0,-0.2) -| node[below,pos=0.25] {Counterexample $\state$} (synthesiser.320);
	\draw[->,thick] (synthesiser.40) -- +(0,0.25) -| node[above,pos=0.25] {$\inv \in \generator{\temp}$}  (verifier.140);
	\draw[->,thick] (tcreator.north) -- +(0,0.4) -| node[above,pos=0.12] {Template $\temp$} (synthesiser);
	\draw[->,thick] (synthesiser.south) -- +(0,-0.4) -| node[below,pos=0.38] {Unsat\gray{, hint?}} (tcreator);
	
\end{tikzpicture}
}
\caption{Our CEGIS framework for synthesizing quantitative inductive invariants.}
\label{fig:framework}
\end{figure}

\smallskip\noindent
4) \emph{Inductive Synthesis.}
Our approach to finding invariants, as sketched in \Cref{fig:framework},
 is inspired by \emph{inductive synthesis}~\cite{DBLP:series/natosec/AlurBDF0JKMMRSSSSTU15}:
The inner loop (shaded box) is provided with a  \emph{template} $T$ which may generate an infinite set $\generator{\temp} $ of instances. 
We then synthesize a template instance $I$ that is 
an inductive invariant witnessing quantitative reachability, or determine that no such instance exists. 
We search for such instances in a \emph{counterexample-guided inductive synthesis} (CEGIS) loop: 
The synthesizer constructs a candidate. (A tailored variant of) an off-the-shelf verifier either
\begin{enumerate*}[label=(\roman*)]
	\item decides that the candidate is a suitable inductive invariant or 
	\item reports a counterexample state $\state$ back to the synthesizer.
\end{enumerate*}
Upon termination (guaranteed for finite-state programs), the inner loop has either found an inductive invariant or the solver reports that the template $\temp$ 
does not admit an inductive invariant. 

\medskip\noindent
\textbf{Contributions.}
We show that inductive synthesis for verifying \emph{quantitative reachability properties} by finding \emph{inductive invariants} on \emph{source-code level} is feasible: Our approach is sound for arbitrary probabilistic programs, and complete for finite-state programs.  We implemented our simple yet powerful technique. The results are promising: Our CEGIS loop is sufficiently fast to support large templates and finds inductive invariants for various probabilistic programs and properties. It can prove, amongst others, upper and lower bounds on reachability probabilities and universal positive almost-termination~\cite{benni_diss}.
Our implementation is competitive with three state-of-the-art tools -- \toolname{Storm}~\cite{DBLP:journals/corr/abs-2002-07080}, \toolname{Absynth}~\cite{DBLP:conf/pldi/NgoC018}, and \toolname{Exist}~\cite{DBLP:conf/cav/BaoTPHR22} --
on subsets of their benchmarks fitting our framework.


\smallskip\noindent
\emph{Applicability and Limitations.}
We consider programs with possibly unbounded nonnegative integer-valued variables and arbitrary affine expressions in quantitative specifications.
As for other synthesis-based approaches, there are unrealizable cases -- loops for which no piecewise linear invariant exists.
But, if there is an invariant, our CEGIS loop often finds it within a few iterations. 

	\section{Overview}
\label{sec:overview}

\begin{figure}[t]
\belowdisplayskip=0pt%
\begin{align*}
		\gray{\mathtt{1}\colon}\quad&\COMPOSE{\COMPOSE{
			\ASSIGN{\varfailed}{0}}
			{\ASSIGN{\varsent}{0}}}{}\\
		\gray{\mathtt{2}\colon}\quad
		&\WHILE{\varsent < 8\,000\,000 \wwedge \varfailed < 10} 	 \\
		\gray{\mathtt{3}\colon\quad}
		& \qquad \PCHOICE{\COMPOSE{\ASSIGN{\varfailed}{0}}{
			\ASSIGN{\varsent}{\varsent + 1}
		} }
		%
		{0.999} 
		%
		{\ASSIGN{\varfailed}{\varfailed + 1}} \quad\}
		%
\end{align*}
\caption{Model for the bounded retransmission protocol (BRP).}
\label{fig:brp}
\end{figure}%
We illustrate our approach using the bounded retransmission protocol (BRP) -- a standard probabilistic model checking benchmark~\cite{DBLP:conf/types/HelminkSV93,DBLP:conf/papm/DArgenioJJL01} -- modeled by the probabilistic program in \Cref{fig:brp}.
    The model attempts to transmit 8 million packets\footnote{Large constants like the number of packets appear naturally in quantitative models of protocols and have a non-trivial impact on probabilities.} over a lossy channel, where each packet is lost with probability $0.1\%$; 
    if a packet is lost, we retry sending it; if any packet is lost in 10 consecutive sending attempts ($\varfailed = 10$), the \emph{entire} transmission fails; if all packets have been transmitted successfully ($\varsent = 8\,000\,000$), the transmission succeeds.

\subsection{Reachability Probabilities and Loops}
\label{sec:overview:reach}

We aim to reason about the transmission-failure probability of~BRP, i.e. the probability that the loop terminates in a target state $\stateB$ with $\stateB(\varfailed) = 10$ when started in initial program state $\initialState$ with \mbox{$\initialState(\varfailed) = \initialState(\varsent) = 0$}. 
One approach to determine this probability is to
\begin{enumerate*}[label=(\roman*)]
	\item construct an explicit-state Markov chain~(MC) per \Cref{fig:brp},
	\item derive its Bellmann operator $\Phi$~\cite{Puterman1994Markov},
	\item compute its least fixed point $\lfp \Phi$ (a vector containing for \underline{\emph{each}} state the probability to reach $t$), e.g.\ using value iteration (\cf~\cite[Thm~10.15]{Baier2008Principles}), and finally
	\item evaluate $\lfp \Phi$ at~$\initialState$.
\end{enumerate*}

The explicit-state MC of BRP has ca.~80 million states.
We \emph{avoid} building such large state spaces by computing a symbolic representation of $\Phi$ from the program.
More formally, 
let $\States$ be the set of all states,
$\mathtt{loop}$ the entire loop (ll.~2--3 in \Cref{fig:brp}),
$\mathtt{body}$ the $\mathtt{loop}$'s body (l. 3),
and $\eval{\mathtt{body}}(\state)(\state')$ the probability of reaching state~$\state'$ by executing $\mathtt{body}$ once on state $\state$.
Then the least fixed point of the $\mathtt{loop}$'s Bellmann operator $\Phi\colon \bigl(\States \to \PosRealsInf \bigr) \to \bigl(\States \to \PosRealsInf\bigr)$, defined by
\begin{align*}
	\Phi(I) \eeq \lambda \state.~  \begin{cases}
		1, & \textnormal{if } \state(\varfailed) = 10~,\\[.5em]
		\displaystyle{\sum_{s' \in \States}} \: \eval{\mathtt{body}}(s)(s') \cdot I(s'),& \!\begin{array}{l}\textnormal{if } \state(\varsent) < 8\,000\,000\\[-.5ex] \quad \textnormal{ and } \state(\varfailed) < 10~,\end{array}\\[1.5em]
		0,& \textnormal{otherwise } ,
	\end{cases}
\end{align*}%
captures the transmission-failure probability for the \emph{entire} execution of $\mathtt{loop}$ and for \emph{any} initial state, that is, $(\lfp \Phi)(\state)$ is the probability of terminating in a target state when executing $\mathtt{loop}$ on $\state$ (even if $\mathtt{loop}$ would not terminate almost-surely).
Intuitively, $\Phi(I)(\state)$ maps to $1$ if $\mathtt{loop}$ has terminated meeting the target condition (transmission failure); and to 0 if $\mathtt{loop}$ has terminated otherwise (transmission
 success).
If $\mathtt{loop}$ is still running (i.e.\ it has neither failed nor succeeded yet), then $\Phi(I)(\state)$ maps to the expected value of $I$ after executing $\mathtt{body}$ on state $\state$.

\subsection{Quantitative Inductive Invariants}
\label{sec:overview:indinvariants}

Reachability probabilities are generally not computable for infinite-state probabilistic programs~\cite{DBLP:journals/acta/KaminskiKM19}. Even for finite-state programs the state-space explosion may prevent us from computing reachability probabilities exactly. 
However, it often suffices to know that the reachability probability is bounded from above by some threshold~$\lambda$.
For BRP, we hence aim to prove that $(\lfp \Phi)(\initialState) \leq \lambda$.

%
%
%
We attack the above task by means of \emph{(quantitative) inductive invariants}:
a~candidate for an inductive invariant is a mapping~$I\colon \States \rightarrow \PosRealsInf$. \emph{Intuitively}, such a candidate $I$~is \emph{inductive} if the following holds:
when assuming that $I(\state)$ is (an over-approximation of) the probability to reach a target state upon termination of $\mathtt{loop}$ on $s$, then the probability to reach a target state after performing one more guarded loop iteration, \ie executing $\ITNE{\varsent < {\ldots}}{\COMPOSE{\mathtt{body}}{\mathtt{loop}}}$ on $\state$, must be \emph{at most} $I(\state)$. \emph{Formally}, $I$ is an inductive invariant\footnote{For an exposition of why it makes sense to speak of \emph{invariants} even in a quantitative setting, \cite[Sect.~5.1]{benni_diss} relates quantitative invariants to invariants in Hoare logic.} if%
\begin{align*}
	\forall \state\colon \quad \Phi(I)(\state) \lleq I(\state) \qquad\textnormal{which implies}\qquad \forall \state\colon \quad  \bigl(\lfp \Phi \bigr) (\state) \lleq I(\state)
\end{align*}
by Park induction~\cite{park1969fixpoint}.
Hence, $I(\state)$ bounds for each initial state $\state$ the exact reachability probability from above.
If we are able to find an inductive $I$ that is below $\lambda$ for the initial state $\initialState$ with $\varfailed = \varsent = 0$, i.e.\ $I(\initialState) \leq \lambda$, then we have indeed proven the upper bound $\lambda$ on the transmission-failure probability of our BRP model.
In a nutshell, our goal can be phrased as follows:

\smallskip
\noindent
\fbox{\begin{minipage}{0.99\textwidth}
 \textbf{Goal:} Find an inductive invariant $\inv$, i.e.\ an $\inv$ with $\Phi(I) \leq I$,
 s.t.\ $I(s_0) \leq \lambda$.
\end{minipage}}

\subsection{Our CEGIS Framework for Synthesizing Inductive Invariants}\label{sec:overviewcegis}
While \emph{finding} a safe inductive invariant $I$ is challenging, 
\emph{checking} whether a given candidate $I$ is indeed inductive is easier: 
it is decidable for certain infinite-state programs (\cf \cite[Sect.~7.2]{DBLP:conf/cav/BatzCKKMS20}), it may not require an explicit exploration of the whole state space, and 
it can be done efficiently for piecewise linear $I$.
Hence, techniques that generate decent candidate expressions fast and then check their inductivity could enable the automatic verification of probabilistic programs with gigantic and even infinite state spaces.

In this paper, we test this hypothesis by developing
the CEGIS framework depicted in \Cref{fig:framework} for incrementally synthesizing inductive invariants.
%
%
A template generator generates parametrized templates for inductive invariants.
The inner loop (shaded box in \Cref{fig:framework}) then tries to solve for appropriate template-parameter instantiations.
If it succeeds, an inductive invariant has been synthesized.
Otherwise, the template provably cannot be instantiated into an inductive invariant.
The inner loop then reports that back to the template generator (possibly with some hint on why it failed, see \Cref{sec:refinement}) and asks for a refined template.

For our running example, we start with the template%
\begin{align}
	T \eeq \iverson{\varfailed<10 \land \varsent < 8\,000\,000} \cdot (\alpha \cdot \varsent + \beta \cdot \varfailed + \gamma) \pplus \iverson{\varfailed=10},
	\label{eq:overview:first-template}	
\end{align}%
where we use \emph{Iverson brackets} for indicators, i.e.\ $\iverson{\varphi}(s) = 1$ if $s \models \varphi$ and $0$ otherwise.
$T$ contains two kinds of variables: integer program variables $\varfailed, \varsent$ and $\Rats$-valued parameters $\alpha, \beta, \gamma$. 
While the template is nonlinear, substituting $\alpha, \beta, \gamma$ with concrete values yields piecewise linear candidate invariants $I$.  We ensure that those $I$ are piecewise linear to render the repeated inductivity checks 
efficient. 
We construct only so-called \emph{natural} templates $T$  with $\Phi$ in mind, e.g.\ we want to construct only $I$ such that $I(s) = 1$ when $s(\varfailed) = 10$.  

Our inner CEGIS loop checks whether there exists an assignment from these template variables to concrete values such that the resulting piecewise linear expression is an inductive invariant. 
%
 %
Concretely, we try to determine  whether there exist values for $\alpha, \beta, \gamma$ such that $T(\alpha, \beta,\gamma)$ is inductive. 
For that, we first guess values for $\alpha, \beta, \gamma$, say all $0$'s, and ask a verifier whether the instantiated (and now piecewise linear) template $I = T(0,0,0)$ is indeed inductive.
In our example, the verifier determines that $I$ is \emph{not} inductive: a counterexample is $\state(\varfailed)=9$, $\state(\varsent)=7999999$.
Intuitively, the probability to reach the target after one more loop iteration exceeds the value in $I$ for this state, that is, $\Phi(I)(\state) = 0.001 > 0 = I(\state)$.
From this counterexample, our synthesizer
learns%
\begin{align*}
	\Phi(T)(s) \eeq 0.001 \smash{~{}\stackrel{!}{\leq}{}~} \alpha \cdot 7999999 + \beta \cdot 9 + \gamma \eeq T(s)~.
\end{align*}%
Observe that this learned lemma is linear in $\alpha, \beta, \gamma$. 
The synthesizer will now keep \enquote{guessing} assignments to the parameters which are consistent with the learned lemmas until either no such parameter assignment exists anymore, or until it produces an \emph{inductive} invariant $I = T(\hdots)$. 
In our running example, assuming $\lambda = 0.9$, after $6$ lemmas, our synthesizer finds \mbox{the inductive invariant $I$}%
%
\begin{align}
	\iverson{\varfailed<10 \land \varsent < 8 \cdot 10^6} \cdot (-\tfrac{9}{8\cdot 10^7} \cdot \varsent + \tfrac{79\,991}{72\cdot 10^7} \cdot \varfailed + \tfrac{9}{10}) + \iverson{\varfailed=10}
	\label{eq:overview:first-invariant}	
\end{align}%
\normalsize%
where indeed $I(s_0) \leq \lambda$ holds.
For a tighter threshold $\lambda$, such simple templates do not suffice. 
For example, it is impossible to instantiate this template to an inductive invariant for $\lambda = 0.8$, even though $0.8$ is an upper bound on the actual reachability probability. 
We therefore support \emph{more general templates} of the form%
\begin{align*}
	T \eeq \sum_{\smash{i}} \iverson{B_i} \cdot \left(\alpha_i \cdot \varsent + \beta_i \cdot \varfailed + \gamma_i \right) \pplus \iverson{\varfailed=10}~,
\end{align*}%
where the $B_i$ are (restricted) predicates over  program and template variables which partition the state space. 
In particular, we allow for a template such as%
	\begin{equation}
	 \begin{aligned} 
	 	T \eeq 
		& \iverson{\varfailed < 10 \land \varsent < \delta} \cdot \left( \alpha_1 \cdot \varsent + \beta_1 \cdot \varfailed + \gamma_1  \right) \pplus  \\
		& \iverson{\varfailed < 10 \land \varsent \geq \delta} \cdot \left( \alpha_2 \cdot \varsent + \beta_2 \cdot \varfailed + \gamma_2 \right) \pplus \iverson{\varfailed=10} 
	\end{aligned}%
	\label{eq:varpartemplexample}
	\end{equation}%
However, such templates are challenging for the CEGIS loop. Thus, we additionally consider templates where the $B_i$'s range only over program variables, \eg
	%
	\small%
	\begin{align*} 
		%
		& \iverson{\varfailed < 10 \land \varsent < 4\,000\,000} \cdot  (\hdots )
		 \pplus 
		%
		\iverson{\varfailed < 10 \land \varsent \geq 4\,000\,000} \cdot (\hdots)
		\pplus  \hdots 
	\end{align*}%
	\normalsize%
Our partition refinement algorithms automatically produce these templates, without the need for user interaction. 

\begin{figure}[t]
\begin{subfigure}{0.65\textwidth}
\begin{align*} 
	 &\COMPOSE{\COMPOSE{
			\ASSIGN{\varfailed}{0}}
		{\ASSIGN{\varsent}{0}}}{}\\
		&\WHILE{\varsent < P \wwedge \varfailed < R \wwedge P \leq 8\,000\,000 \wwedge R \geq 5} 	 \\
		& \hspace{1em} \PCHOICE{\COMPOSE{\ASSIGN{\varfailed}{0}}{
			\ASSIGN{\varsent}{\varsent + 1}
		} }
		%
		{0.99} 
		%
		{\ASSIGN{\varfailed}{\varfailed + 1}}  \\
		&\}	
\end{align*}
\caption{A family of retransmission protocols}
\label{fig:brpfam}
\end{subfigure}
\begin{subfigure}{0.3\textwidth}
		\includegraphics[width=0.99\textwidth]{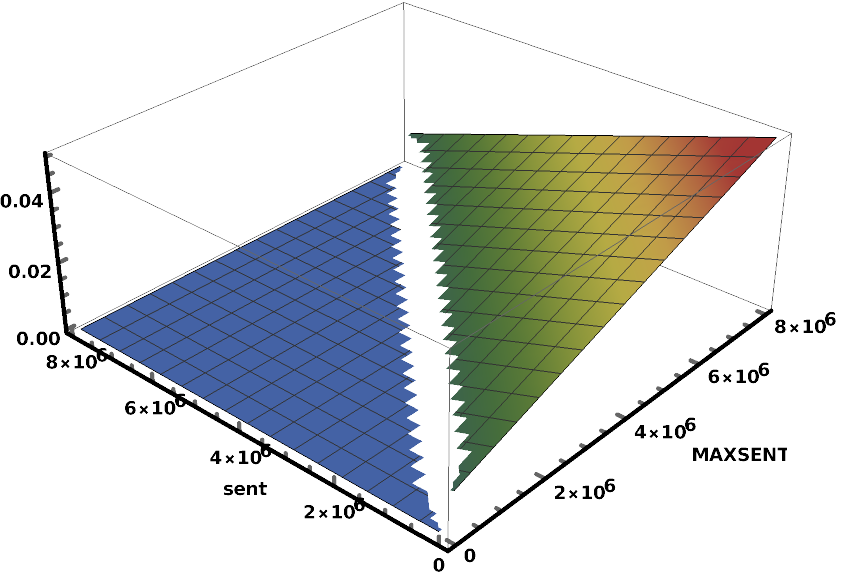}
	\caption{Inductive invariant for $\varfailed=0$ and $R \geq 5$}
	\label{fig:invplot}
\end{subfigure}
\caption{A bounded retransmission protocol family and piece of a matching invariant.}
\end{figure}

Finally, we highlight that we may use our approach for more general questions. 
For BRP, suppose we want to verify an upper bound $\lambda = 0.05$ on the probability of failing to transmit \emph{all} packages for an \emph{infinite set of models} (also called a \emph{family}) with varying upper bounds on packets $1 \leq P \leq 8000000$ and retransmissions $R \geq 5$. This infinite set of models is described by the loop shown in \Cref{fig:brpfam}. Our approach
fully automatically synthesizes the following inductive invariant $\inv$:
\small%
\begin{align*}
    \label{eq:overview:final}
    	& \left[ \begin{array}{l} 
		\varfailed < R \wwedge \varsent<P\wwedge P<8\,000\,000 \wwedge R\geq5  \\[1ex]
		\wwedge R > 1+\varfailed \wwedge \tfrac{13067990199}{5280132671650}\cdot\varfailed \leq \tfrac{5278689867}{211205306866000}
	\end{array} \right]\cdot 
	\left(\begin{array}{l}
		\tfrac{-19}{3820000040}\cdot \varsent \\[1ex]
		{}+ \tfrac{19}{3820000040} \cdot P \\[1ex]
		{}+ \tfrac{19500001}{1910000020}
	\end{array}\right) \\[-1ex]
	%
	%
	\pplus & \ldots~ \textnormal{(7 additional summands omitted)}
\end{align*}%
\normalsize%
%
%
%
%
The first summand of $\inv$ is plotted in \Cref{fig:invplot}. Since $\inv$ overapproximates the probability of failing to transmit all packages for every state, $\inv$ may be used to infer additional information about the reachability probabilities. 

%
%
%
	\section{Formal Problem Statement}	\label{sec:problem-statement}

Before we state the precise invariant synthesis problem that we aim to solve, we summarize the essential concepts underlying our formalization.

\smallskip\noindent%
\textit{Probabilistic Loops.}
We consider \emph{single probabilistic loops}
$\WHILEDO{\Guard}{\Body}$ whose \emph{loop guard} $\Guard$
and (loop-free) \emph{body} $\Body$ adhere to the grammar
\begin{align*}
     \cc ~~{}\longrightarrow{}~~ 
	 &\SKIP	~\mid~ \ASSIGN{x}{\ee}  ~\mid~  \COMPOSE{\cc}{\cc} 
     ~\mid~  \PCHOICE{\cc}{p}{\cc} 
	%
    ~\mid~ \ITE{\guard}{\cc}{\cc} \\
    \guard ~~{}\longrightarrow{}~~ 
    & \ee < \ee ~\mid~ \neg \guard ~\mid~ 
      \guard \wedge \guard 
    \qquad\quad
    \ee ~~{}\longrightarrow{}~~ 
    z ~\mid~ x ~\mid~ z \cdot \ee ~\mid~ \ee + \ee ~, 
    %
	%
\end{align*}%
where $z \in \Ints$ is a constant and $x$ is from
an arbitrary \emph{finite} set $\Vars$ of $\Nats$-valued program variables. Program states in $\States = \setcomp{ \state}{\state\colon \Vars \to \Nats }$
map variables to natural numbers.\footnote{%
Considering only unsigned integers does not decrease expressive power but simplifies the technical presentation (\cf \cite[Sect.~11.2]{relatively_complete_prob_progs} for a detailed discussion).
We statically ensure that for every assignment $\ASSIGN{x}{e}$, $e$ always evaluates to some value in $\Nats$. 
}
All statements are standard (\cf \cite{McIverM05}). 
$\PCHOICE{\cc_1}{p}{\cc_2}$ is a probabilistic choice which executes $\cc_1$ with probability \mbox{$p \in [0,1] \cap \Rats$} and $\cc_2$ with probability $1-p$. 
\Cref{fig:brp} (ll.\ 2--3) is an example of a probabilistic loop.

\smallskip\noindent%
\textit{Expectations.}
In \Cref{sec:overview}, we considered whether final states meet some target condition by assigning $0$ or $1$ to each final state.
The assignment can be generalized to more general 
quantities in $\PosRealsInf$. 
We call such assignments $f$ \emph{expectations}~\cite{McIverM05} (think: random variable)
and collect them in the set $\E$, \ie 
\begin{align*}
	\E \eeq \setcomp{f}{f \colon \States \to \PosRealsInf}~,\qquad \!\textnormal{where}\!\qquad f \ppreceq g \qiff \forall\, s \in \States\colon f(s) \leq g(s)~.
\end{align*}%
$\preceq$ is a partial order on $\E$ -- necessary to sensibly speak about least fixed points.

\smallskip\noindent%
\textit{Characteristic Functions.}
The expected behavior of a probabilistic loop for an expectation $f$ is captured by an expectation transformer (namely the $\Phi\colon \E \to \E$ of \Cref{sec:overview}), 
called the loop's \emph{characteristic function}.
To focus on invariant synthesis, we abstract from the 
details\footnote{%
We can (and our tool does) derive a symbolic representation of a loop's characteristic function from the program structure using a weakest-precondition-style calculus (\cf \cite{McIverM05}); see \Cref{app:wp} for details. 
If $f$ maps only to $0$ or $1$, $\charfun{f}$ corresponds to the least fixed point characterization of reachability probabilities~\cite[Thm.~10.15]{Baier2008Principles}.%
} %
of constructing characteristic functions from probabilistic loops; our framework only requires the following key property:%
\begin{proposition}[Characteristic Functions]\label{prop:charfun}
	For every loop $\WHILEDO{\Guard}{\Body}$ and expectation $f$, there exists a monotone function
	 $\charfun{f}\colon \E \to \E$ such that%
		\begin{align*}
			\charfun{f}(I)(s) \eeq \begin{cases}
				f(s), & \textnormal{if } s \not\models \Guard~,\\[.5em]
				\textnormal{\enquote{expected value of $I$ after executing $\Body$ \emph{once} on $s$}},& \textnormal{if } s \models \Guard~,
		\end{cases}
	\end{align*}%
	and the least fixed point of $\charfun{f}$, denoted $\lfp \charfun{f}$, satisfies%
	\begin{align*}
		\bigl(\lfp \charfun{f}\bigr)(s) \eeq \textnormal{\enquote{expected value of $f$ after executing $\WHILEDO{\Guard}{\Body}$ on $s$}}~.
	\end{align*}
\end{proposition}%
\begin{example}\label{ex:running}
In our running example from \Cref{sec:overview:reach}, 
we chose as $f$ the expression $\iverson{\varfailed = 10}$, which evaluates to $1$ in every state $s$ where $\varfailed = 10$ and to $0$ otherwise.
The characteristic function $\charfun{f}(I)$ of the loop
in \Cref{fig:brp} is  

\noindent$
	\iverson{\neg \varphi} \cdot \iverson{\varfailed {=} 10} \pplus
	\iverson{\varphi} \cdot \bigl(0.999 \cdot I\subst{\varsent}{\varsent{+}1}\subst{\varfailed}{0} + 0.001 \cdot I\subst{\varfailed}{\varfailed {+} 1}\bigr)
$,

\noindent
where $\Guard = \varsent < 8\,000\,000 \wedge \varfailed < 10$ is the loop guard and
$I\subst{x}{\ee}$ denotes the (syntactic) substitution of variable $x$ by expression $\ee$ in expectation $I$ -- the latter is used to model the effect of assignments as in standard Hoare logic.
%
\qedTT
\end{example}

\smallskip\noindent%
\textit{Inductive Invariants.}
For a probabilistic loop $\WHILEDO{\Guard}{\Body}$, and \emph{pre}- and \emph{post}expectations $g,f \in \E$,
we aim to verify $\lfp \charfun{f} \preceq g$, i.e.\ that the expected value of~$f$ after termination of the loop is bounded from above by $g$. We discuss how to adapt our approach to expected runtimes and lower bounds in \Cref{sec:past_lower}.
Intuitively, $f$ assigns a quantity to all \emph{target} states reached upon termination.
$g$ assigns to all \emph{initial states} a desired bound on the expected value of $f$ after termination of the loop.
By choosing $g(\state) = \infty$ for certain $s$, we can make $\state$ so-to-speak \enquote{irrelevant}.
An $I \in \E$ is an \emph{inductive invariant} 
proving $\lfp \charfun{f} \preceq g$ 
iff 
$\charfun{f}(I) \preceq I$ and $I \preceq g$.
Continuing our 
example, \Cref{eq:overview:first-invariant} on p.~\pageref{eq:overview:first-invariant} shows an inductive invariant proving that $ \lfp \charfun{f} \preceq g \coloneqq [\varfailed=0 \wedge \varsent=0]\cdot 0.9 + [\neg(\varfailed=0 \wedge \varsent=0)]\cdot \infty$.

Our framework employs syntactic fragments of expectations on which the check $\charfun{f}(I) \preceq I$ can be done
symbolically by an SMT solver. 
As illustrated in \Cref{fig:framework}, we use \emph{templates} to further narrow down the invariant search space.

\smallskip\noindent%
\textit{Templates.}
Let $\tvars = \{\alpha, \beta, \ldots \}$ be a countably infinite set of $\Rats$-valued \emph{template variables}.
A \emph{template valuation} is a function $\tval\colon \tvars \to \Rats$ that assigns to each template variable a rational number.
We will use the same expressions as in our programs except that we admit both rationals and template variables as coefficients.
Formally, 
arithmetic and Boolean expressions $\Te$ and $\Tb$ adhere to%
%
\begin{align*}
	\Te \quad{}&\longrightarrow{}\quad   \rr \mmid x \mmid \rr \cdot x \mmid \Te + \Te \qquad\quad\:
	\Tb \quad{}\longrightarrow{}\quad \Te < \Te \mmid~ \neg \Tb \mmid~ \Tb \wedge \Tb~,
\end{align*}%
where $x \in \Vars$ and $\rr \in \Rats \cup \tvars$.
The set $\Temp$ of templates then consists of all%
\begin{align*}
	\temp \eeq 
	\iverson{\Tb_1} \cdot \Te_1
	+ \ldots +
	\iverson{\Tb_n} \cdot \Te_n~,
\end{align*}%
for $n \geq 1$, where \emph{the Boolean expressions $\Tb_i$ partition the state space}, \ie for all template valuations $\tval$ and all states $\state$, there is \emph{exactly one} $\Tb_i$ such that $\tval, \state \models \Tb_i$.
$\temp$~is a \emph{fixed-partition template} if additionally no $\Tb_i$ contains a template variable.

Notice that templates are generally \emph{not} linear (over $\Vars \cup \tvars$).
\Cref{sec:overview} gives several examples of templates, \eg \Cref{eq:overview:first-template}.

\smallskip\noindent%
\textit{Template Instances.}
We denote by $\tvalapps{\temp}$ the \emph{instance} of template $\temp$ under $\tval$, \ie the expression obtained from substituting every template variable $\alpha$ in $\temp$ by its valuation $\tval(\alpha)$.
For example, the expression in \Cref{eq:overview:first-invariant} on p.~\pageref{eq:overview:first-invariant} is an instance of the template in \Cref{eq:overview:first-template} on p.~\pageref{eq:overview:first-template}.
The set of all instances of template $\temp$ is defined as
$
\generator{\temp} = \setcomp{ \tvalapps{\temp} }{ \tval \colon \tvars \to \Rats }
$.
We chose the shape of templates on purpose:
To evaluate an instance $\tvalapps{\temp}$ of a template $\temp$ in a state $\state$, it suffices to find the \emph{unique} Boolean expression $\Tb_i$ with $\tval, \state \models \Tb_i$ and then evaluate the \emph{single} linear arithmetic expression $\tvalapps{\Te_i}$ in $\state$.
For fixed-partition templates, the selection of the right $\Tb_i$ does not even depend on the template evaluation $\tval$.

\smallskip\noindent%
\textit{Piecewise Linear Expectations.}
Some template instances $\tvalapps{\temp}$ do \emph{not} represent expectations, \ie they are not of type $\States \to \PosRealsInf$, as they may evaluate to \emph{negative numbers}.
Template instances $\tvalapps{\temp}$ that \emph{do} represent expectations are \emph{piecewise linear}; we collect such \emph{well-defined} instances in the set $\Elin$. Formally,%
\begin{definition}[$\boldsymbol{\Elin}$]
  The set $\Elin$ of \emph{(piecewise) linear expectations}
  is %
  \mbox{$
     \Elin \eeq \{ \tvalapps{\temp} \mid \temp \in \Temp ~~\textnormal{and}~~ \tval\colon \tvars \to \Rats ~~\textnormal{and}~~ \forall \state \in \States\colon \tvalapps{\temp}(\state) \geq 0 \}
  $}.
\end{definition}%
We identify well-defined instances of templates in $\Elin$ with the expectation in~$\E$ that they represent, \eg when writing the inductivity check $\charfun{f}(\tvalapps{\temp}) \stackrel{\smash{\raisebox{-.5ex}{\textnormal{\tiny ?}}}}{\preceq} (\tvalapps{\temp})$.

\smallskip\noindent%
\textit{Natural Templates.}
As suggested in \Cref{sec:overviewcegis}, it makes sense to focus only on so-called \emph{natural} templates.
Those are templates that even have a chance of becoming inductive, as they take the loop guard $\Guard$ and postexpectation~$f$ into account. 
%
%
Formally, a template $\temp$ is \emph{natural} (wrt. to $\Guard$ and $f$) if $\temp$ is of the form%
\begin{align*}
		\temp ~\eeq~ \underbrace{\iverson{\neg \Guard \wedge \Tb_1} \cdot \Te_1
	+ \ldots +
	\iverson{\neg \Guard \wedge \Tb_n} \cdot \Te_n}_{\textnormal{must be equivalent to $\iverson{\neg\Guard} \cdot f$}}
	\pplus \iverson{\Tb'_1} \cdot \Te'_1
	+ \ldots +
	\iverson{\Tb'_m} \cdot \Te'_m~.
\end{align*}%
%
We collect all natural templates in the set $\Ntemp$.

\medskip\noindent%
\textbf{Formal Problem Statement.}\label{problem-statement}
Throughout this paper, we fix an ambient single loop
$\WHILEDO{\Guard}{\Body}$, a postexpectation $f \in \Elin$, and
a preexpectation $g  \in \Elin$\footnote{To enable declaring certain states as irrelevant, we additionally allow $\Te_i = \infty$ in the linear preexpectation $g = \iverson{\Tb_1} \cdot \Te_1 + \ldots + \iverson{\Tb_n} \cdot \Te_n$.} such that \mbox{$\lfp \charfun{f}(I) \preceq g$}\footnote{We discuss in \Cref{sec:past_lower} how to reason about lower bounds $g \preceq \lfp \charfun{f}(I)$.}.
The set $\indinvshort$ of \emph{admissible invariants} (\ie those expectations that are both \emph{inductive} and \emph{safe}) is then given by%
%
\begin{align*}
   \indinvshort \eeq \{ \:\:\underbrace{~I \iin \Elin~}_{\mathclap{\text{well-definedness:}~I\succeq 0}} \qmid \underbrace{~\charfun{f}(I) \ppreceq I~}_{\text{inductivity}} \qand \underbrace{~I \ppreceq g~}_{\text{safety}} \:\:\}~,
\end{align*}%
where the underbraces summarize the tasks for a verifier to decide whether a template instance $I$ is an admissible inductive invariant.
%
We require \mbox{$\lfp \charfun{f} \preceq g$}, so that $\indinvshort$ is not vacuously empty due to an unsafe bound $g$.

\begin{mdframed}
	 \textbf{Formal problem statement}: Given a natural template $\temp$, find an instantiation $\inv \in \generator{\temp} \cap \indinvshort$ or determine that there is no such $\inv$.
\end{mdframed}
Notice that $\indinvshort$ might be empty, even for safe $g$'s, because generally one might need more complex invariants than piecewise linear ones~\cite{relatively_complete_prob_progs}.
However, there always exists an inductive invariant in $\Elin$
if a loop can reach only finitely many states.\footnote{Bluntly just choose as many pieces as there are states.}
We call a loop $\WHILEDO{\guard}{\cc}$ \emph{finite-state}, if only finitely many states satisfy the loop guard $\Guard$, i.e.\  if
$\States_\Guard = \setcomp{\state \in \States }{ \state \models \Guard}$ is finite.

\medskip\noindent%
\textbf{Syntactic Characteristic Functions.}
We work with \emph{linear} expectations \mbox{$I,f \in \Elin$}, so that we can check inductivity ($\charfun{f}(I) \preceq I$) symbolically (via SMT) without state space construction.
In particular, we can construct a \emph{syntactic counterpart} $\tcharfun{f}$ to $\charfun{f}$ that operates on \emph{templates}.
%
Intuitively, whether we evaluate $\tcharfun{f}$ on a (syntactic) template $\temp$ and then instantiate the result with a valuation $\tval$, \emph{or} we evaluate $\charfun{f}$ on the (semantic) expectation $\tvalapps{\temp}$ emerging from instantiating $\temp$ with $\tval$ -- the results will coincide if $\tvalapps{\temp}$ is well-defined.
Formally:%
\begin{proposition}
	\label{thm:templatetrans}
	Given $\WHILEDO{\guard}{\cc}$ and
	$f \in \Elin$, one can effectively compute a mapping
	$\tcharfun{f} \colon \Temp \to \Temp$, such that for all $\temp$\! and $\tval$
	\[
	\tvalapps{\temp} \iin \Elin
	\qimplies
	\tvalapps{\tcharfun{f}(\temp)} \eeq \charfun{f}\bigl( \tvalapps{\temp} \bigr)~.
	\]
	Moreover, $\tcharfun{f}$ maps fixed-partition templates to fixed-partition templates.
\end{proposition}%
%
%
%
\noindent%
In \Cref{ex:running}, we have already constructed such a $\tcharfun{f}$ 
to represent $\charfun{f}$. 
The general construction is inspired by \cite{DBLP:conf/cav/BatzCKKMS20}, but treats template variables as constants.

	\section{One-Shot Solver}
	\label{sec:oneshot}

\noindent
One could address the template instantiation problem from \Cref{sec:problem-statement} in one shot: 
encode it as an SMT query, ask a solver for a model, and infer from the model an admissible invariant.
While this approach is infeasible in practice (as it involves quantification over $\States_\Guard$), it inspires the CEGIS loop in \Cref{fig:framework}.

Regarding the encoding, given a template $\temp$, we need a formula over $\tvars$ that is satisfiable if and only if there exists a template valuation $\tval$ such that
$\tvalapps{\temp}$ is an admissible invariant, \ie $\tvalapps{\temp} \in \indinvshort$.
To get rid of program variables in templates, we denote by $\temp(\state)$ the expression over $\tvars$ in which all \emph{program} variables $x \in \Vars$ have been substituted by $\state(x)$.

Intuitively, we then encode that, for every state $\state$, the expression $\temp(\state)$ 
satisfies the three conditions of admissible invariants, \ie well-definedness, inductivity, and safety.
In particular, we use \Cref{thm:templatetrans} to compute a template $\tcharfun{f}(\temp)$ that represents the application of the characteristic function $\charfun{f}$ to a candidate invariant, \ie $\charfun{f}(\tvalapps{\temp})$ -- a necessity for encoding inductivity.

Formally, we denote by $\satset(\phi)$ the set of all models of a first-order formula $\phi$ (with a fixed underlying structure), \ie 
$\satset(\phi) = \{ \tval \mid \tval \models \phi \}$. Then:
%
%
%
%
%
%
\begin{theorem}
\label{thm:satforindinv}
	For every natural
	template $\temp \in \Ntemp$ and $f, g \in \Elin$, we have%
	\begin{align*}
	&\generator{\temp} \cap \indinvshort \neq \emptyset\\
	\textnormal{iff} \quad &
	 \satset\big(\:\forall\state \in \States_\Guard \colon ~~
	\underbrace{0 \leq \temp(\state)}_{\mathclap{\text{well-definedness}}} \wwedge \underbrace{\tcharfun{f}(\temp)(\state) \leq \temp(\state)}_{\text{inductivity}} \wwedge \underbrace{\temp(\state) \leq g(\state)}_{\text{safety}}\:\big)
	~\neq~ \emptyset~. 
	\end{align*}%
\end{theorem}%
%
%
Notice that, for fixed-partition templates, the above encoding is particularly simple: $\temp(\state)$ and $\tcharfun{f}(\temp)(\state)$ are equivalent to single linear arithmetic expressions over $\tvars$; $g(\state)$ is either a single expression or $\infty$ -- in the latter case, we get an equisatisfiable formula by dropping the always-satisfied constraint $\temp(s) \leq g(\state)$. 

For general templates, one can exploit the partitioning to break it down into multiple inequalities, \ie every inequality becomes a conjunction over implications of linear inequalities over the template variables $\tvars$.%
\begin{example}
	Reconsider template $\temp$ in \cref{eq:varpartemplexample} on p.~\pageref{eq:varpartemplexample} 
	  %
	  %
	  and assume a state $\state$ with $\state(\varfailed) = 5$ and $\state(\varsent) = 2$. 
	  Then, we encode the well-definedness, $\temp(\state) \geq 0$,
	  as
	  {\footnotesize%
	  \[%
	             \big(5<10 \wedge 2 < \delta \Rightarrow \alpha_1 \cdot 2 + \beta_1 \cdot 5 + \gamma_1 \geq 0\big) 
	             \wedge \big(
	              5<10 \wedge 2 \geq \delta \Rightarrow \alpha_2 \cdot 2 + \beta_2 \cdot 5 + \gamma_2 \geq 0\big)
	  \]%
	  }%
	  where the trivially satisfiable conjunct $5 = 10 \Rightarrow \true$ encoding the last summand, \ie $\iverson{\varfailed = 10}$, has been dropped.
	  \qedTT
\end{example}%
The query in \Cref{thm:satforindinv} involves (non-linear) mixed real and integer arithmetic with quantifiers -- a theory that is undecidable in general.
However, for finite-state loops and natural templates, one can replace the universal quantifier $\forall \state$ by a finite conjunction $\bigwedge_{\state \in \States_\Guard}$ to obtain a (decidable) \texttt{QF\_LRA} formula.%
%
%
%
\begin{theorem}\label{thm:one-shot-finite}
    The problem \mbox{$\generator{\temp} \cap \indinvshort \stackrel{\smash{\raisebox{-.4ex}{\textnormal{\tiny ?}}}}{\neq} \emptyset$} is decidable for finite-state loops and $\temp \in \Ntemp$.
    If \,$\temp$ is fixed-partition, it is decidable via linear programming. 
\end{theorem}
%
%

%
	\section{Constructing an Efficient CEGIS Loop}

\label{sec:fixedtemplsynt}
We now present a CEGIS loop (see inner loop of \Cref{fig:framework}) in which a \emph{synthesizer} and a \emph{verifier} attempt to incrementally solve our problem statement (\cf p.~\pageref{problem-statement}).
%

\subsection{The Verifier}
We assume a verifier for checking $I \stackrel{\smash{\raisebox{-.125ex}{\textnormal{\tiny ?}}}}{\in} \indinvshort$. 
For CEGIS, it is important to get some feedback whenever $I \not\in \indinvshort$.  
To this end, we define:%
%
%
%
%
%
%
%
%
%
%
%
%
\begin{definition}\label{def:partiallyvalid}
    For a state $\state \in \States$, the set $\indinvshort(\state)$ of \emph{$\state$-admissible invariants} is 
    \begin{align*}
    	\indinvshort(\state) \eeq \{\: I ~|~ \underbrace{I(\state) \geq 0}_{\mathclap{\text{$s$-well-defined}}} \qand \underbrace{\charfun{f}(I)(\state) \leq I(\state)}_{\text{$s$-inductive}} \qand \underbrace{I(s) \leq g(s)}_{\text{$s$-safe}} \:\}~.
    \end{align*}%
    For a subset $\States' \subseteq \States$ of states, we define $\indinvshort(\States') = \bigcap_{s \in \States'} \indinvshort(\state)$.
\end{definition}%
%
%
Clearly, 
if $I \not\in \indinvshort$, then 
$I \notin \indinvshort(s)$ for some  $s \in \States$, \ie state $s$ is a \emph{counterexample} to well-definedness, inductivity, or safety of $I$. We denote the set of all such counterexamples (to the claim $I \in \indinvshort$) by $\counterex{\inv}$.
We assume an effective (baseline) verifier for detecting counterexamples:%
%
%
\begin{definition}\label{def:verifier}
	A verifier is \emph{any} function $\verify \colon \Elin \to \{\true\} \cup \States$ such that
	\begin{enumerate}
		\item $\verify(\inv)=\true$ if and only if $\inv \in \indinvshort$, and
		\item $\verify(\inv)=s$ implies $s \in \counterex{\inv}$.
	\end{enumerate}
\end{definition}%
\begin{proposition}[\textnormal{\cite{DBLP:conf/cav/BatzCKKMS20}}]
	There exist effective verifiers.
\end{proposition}%
For example, one can implement an SMT-backed verifier using an encoding analogous to \Cref{thm:satforindinv}, where every model is a counterexample $s \in \counterex{\inv}$:%
	\belowdisplayskip=0pt%
	\begin{align*}
	& I \notin \indinvshort 
	\qiff
	 \underbrace{\satset\Big(\:
	\neg\big(\,0 \leq I \wwedge \charfun{f}(I) \leq I \wwedge I \leq g\,\big)\:\Big)
	\nneq \emptyset}_{\smash{\exists s \in \States\colon ~ I \notin \indinvshort(\state)}}~. 
	\end{align*}%
	\normalsize%
%

\subsection{The Counterexample-Guided Inductive Synthesizer}\label{subsec:synthesizer}
A synthesizer must generate from a given template~$\temp$ instances $I \in \generator{\temp}$ which can be passed to a verifier for checking admissibility.
To make an informed guess, our synthesizers can take a finite set of witnesses $\States' \subseteq \States$ into account:%
%
\begin{definition}
Let $\mathsf{FinStates}$ be the set of finite sets of states. 
A \emph{synthesizer} for template $\temp \in \Ntemp$ is \emph{any} function $\tsynt_\temp\colon \mathsf{FinStates} \rightarrow 
\generator{\temp}
 \cup \{ \false \} $ such that
 \begin{enumerate}
	\item if $\tsynt_\temp(\States') = I$, then $I \in\generator{\temp} \cap \indinvshort(\States')$, and
	\item $\tsynt_\temp(\States') = \false$ if and only if $\generator{\temp} \cap\indinvshort(\States') = \emptyset$.
\end{enumerate}
\end{definition}%
%
%
To build a synthesizer
$\tsynt_\temp(\States')$ for finite sets of states $\States' \subseteq \States$, 
we proceed analogously to one-shot solving for finite-state loops (\Cref{thm:one-shot-finite}), \ie we exploit
\begin{align*}
  \tvalapps{\temp} \in \indinvshort(\States')
 ~~\text{iff}~~ \tval \models 
\bigwedge_{\mathclap{\state \in \States'}}\, 
    \underbrace{0 \leq \temp(\state) \wedge \tcharfun{f}(\temp)(\state) \leq \temp(\state) \wedge \temp(\state) \leq g(\state)}_{ \tvalapps{\temp} \,\in\, \indinvshort(s) }~.
\end{align*}%
That is, our synthesizer may return any model $\tval$ of the above constraint system; it can be implemented as one SMT query.
In particular, one can efficiently find such an $\tval$ for fixed-partition templates via linear programming.%
%
%
\begin{theorem}[Synthesizer Completeness]
For finite-state loops and natural templates $\temp \in \Ntemp$, we have $\tsynt_\temp(\States_\Guard) \in \indinvshort$ or \mbox{$\generator{\temp}\cap\indinvshort = \emptyset$.} 
\end{theorem}%
Using the synthesizer and verifier in concert is then intuitive as in \Cref{alg:k_induction}. We incrementally ask our synthesizer to provide a candidate invariant $I$ that is $s$-admissible for all states $s \in S'$. Unless the synthesizer returns $\false$, we ask the verifier whether $I$ is admissible. If yes, we return $I$; otherwise, we get a counterexample $s$ and add it to $S'$ before synthesizing the next candidate. 


	\begin{algorithm}[t]
	\caption{Template-Instance Synthesizer for template $\temp$}
	\label{alg:k_induction}
	\SetKwInput{Input}{input}\SetKwInOut{Output}{output}\SetNoFillComment
	%
	$\synthcex \leftarrow \emptyset$ \;
	%
	\While{$\tsynt_\temp(\synthcex) \neq \false$}{
		$\inv \leftarrow \tsynt_\temp(\synthcex)$ \;
		$\mathit{result} \leftarrow \verify(\inv)$ \;
		\If{$\mathit{result} = \true$} 
		{\Return $\inv$ \tcc*{Verifier returns $\true$, we have $\inv \in \indinvshort$}}
		%
		  $\synthcex \leftarrow \synthcex \cup \{ \mathit{result} \}$ {\tcc*{ $result$ is a counterexample}}

	}
	\Return $\false$ 		\tcc*{$\generator{\temp} \cap \indinvshort = \emptyset$}
\end{algorithm}

\begin{remark}
\label{rem:coop_verif}	
Without further restrictions, the verifier of \Cref{def:verifier} may go into a \emph{counterexample enumeration spiral}.
In \Cref{app:cooperative}, we therefore discuss additional constraints that make this verifier act more cooperatively.
\qedTT
\end{remark}

	\section{Generalization to Termination and Lower Bounds}\label{sec:past_lower}
We extend our approach to 
\begin{enumerate*}[label=(\roman*)]
\item proving \emph{universal positive almost-sure termination} (UPAST) -- termination in finite expected runtime on all inputs, see~\cite[Sect.~6]{benni_diss} -- by synthesizing piecewise linear upper bounds on expected runtimes, and to
\item verifying \emph{lower bounds} on possibly unbounded expected values. 
\end{enumerate*}

\medskip\noindent%
\textbf{UPAST.}
We leverage Kaminski et al.'s weakest-precondition-style calculus for reasoning about expected runtimes \cite{ert,ert_journal}:%
\begin{proposition}\label{prop:ertcharfun}
	For every loop $\WHILEDO{\Guard}{\Body}$, the monotone function
	\begin{align*}
	\ertcharfun\colon\quad\E \to \E, \qquad \ertcharfun(I)(s) \eeq 1 + \charfun{0}(I)(s)~,
	\end{align*}
	obtained from $\charfun{0}$ (\cf \textnormal{\Cref{prop:charfun}}) satisfies\\
	\vspace*{-\baselineskip}
	\vspace*{-.02cm}
	\begin{align*}
	\bigl(\lfp \ertcharfun\bigr)(s) \eeq \begin{array}{l}\textnormal{``expected number of loop guard evaluations}\\\hspace*{4em}\textnormal{when executing $\WHILEDO{\Guard}{\Body}$ on $s$"~.}\end{array}
	\end{align*}%
	\vspace*{-\baselineskip}
	\vspace*{-.02cm}
\end{proposition}%
All properties of $\charfun{0}$ relevant to our approach carry over to $\ertcharfun$, thus enabling the synthesis of inductive invariants $\inv \in \Elin$ satisfying $0 \preceq \inv$ and $\ertcharfun(\inv) \preceq \inv$. Such~$\inv$ \emph{upper-bound the expected number of loop iterations} \cite{ert} and, since expectations in $\Elin$ never evaluate to infinity, 
$\inv$ witnesses UPAST of the $\WHILESYMBOL$-loop.

\medskip\noindent%
\textbf{Lower Bounds.}
Consider the problem of verifying a lower bound $g \preceq \lfp \charfun{f}$ for some loop $\cc' = \WHILEDO{\Guard}{\Body}$. It is straightforward to modify our CEGIS approach for synthesizing \emph{\underline{sub}-invariants}, \ie $\inv \in \Elin$ with $\inv \preceq \charfun{f}(\inv)$. However,  Hark et al.\ \cite{aiminglow} showed that sub-invariants \emph{do not necessarily lower-bound $\lfp \charfun{f}$}; they hence proposed a more involved yet sound induction rule for lower bounds:
%
\begin{theorem}[Adapted from Hark et al.~\textnormal{\cite{aiminglow}}]\label{thm:lowerbounds}
Let $\temp$ be a natural template and $\inv \in \generator{\temp}$. If \,$0\preceq \inv$, $\inv \preceq \charfun{f}(\inv)$, and $\cc'$ is UPAST, then
\begin{align*}
   \underbrace{\exists\, c \in \PosReals ~ \forall\, \state \in \States_\Guard \colon \quad  \charfun{f}\bigl(\abs{\inv -\inv(s)}\bigr)(s) \lleq c}_{\mathclap{\inv~\text{is \emph{conditionally difference bounded (c.d.b.)}}}} \qqimplies \inv \ppreceq \lfp \charfun{f}~.
\end{align*}%
\end{theorem} 
Akin to \Cref{thm:templatetrans}, given $\temp \in \Ntemp$, we can \emph{compute} $\temp' \in \Ntemp$ s.t.\ for all $\tval$,
\begin{align*}
\tvalapps{\temp} \iin \Elin
\qimplies
\tvalapps{\temp'} \eeq \lambda \state \mydot \charfun{f}\bigl(\abs{\tvalapps{\temp} -\tvalapps{\temp}(s)}\bigr)(s)~,
\end{align*}%
which facilitates the extension of our verifier and synthesizer (see \cref{sec:fixedtemplsynt}) for encoding and checking conditional difference boundedness. Hence, we can employ our CEGIS framework for verifying $g \preceq \lfp \charfun{f}$ by
\begin{enumerate*}[label=(\roman*)]
\item proving UPAST of $\cc'$ as demonstrated above and
\item synthesizing a c.d.b.\ sub-invariant $\inv$ with $g \preceq \inv$.
\end{enumerate*}

	\section{Empirical Evaluation}\label{sec:empirical}
%
\noindent
%
We have implemented a prototype of our techniques called \tool\ifblindreview\else\footnote{\faGithub~\url{https://github.com/moves-rwth/cegispro2}}\fi: CEGIS for PRObabilistic PROgrams. The tool is written in Python using 
 pySMT~\cite{pysmt2015} with Z3~\cite{z3} as the backend for SMT solving. 
 \tool proves upper- or lower bounds on expected outcomes of a probabilistic program 
 by synthesizing quantitative inductive invariants.
We investigate the applicability and scalability of our approach with a focus on the expressiveness of piecewise linear invariants. Moreover, we compare with three state-of-the-art tools -- \toolname{Storm} \cite{DBLP:journals/corr/abs-2002-07080}, \toolname{Absynth} \cite{DBLP:conf/pldi/NgoC018}, and \toolname{Exist} \cite{DBLP:conf/cav/BaoTPHR22} --
on subsets of their benchmarks fitting into our framework.
%
%
%
%
%

\smallskip\noindent%
\emph{Template Refinement.}
We start with a fixed-partition template $\temp_1$ constructed automatically from the syntactic structure of the given loop (\ie the loop guard and branches in the loop body, see \eg \cref{eq:overview:first-template}).
If we learn that $\temp_1$ admits no admissible invariant, we generate a refined template $\temp_2$, and so on, until we  find a template $\temp_{i}$ with $\generator{\temp_i}\cap\indinvshort \neq \emptyset$ or realize that no further refinement is possible.
We implemented three strategies for template refinement (including one producing non-fixed-partition templates); see \Cref{sec:refinement} for details.


\begin{figure}[t]
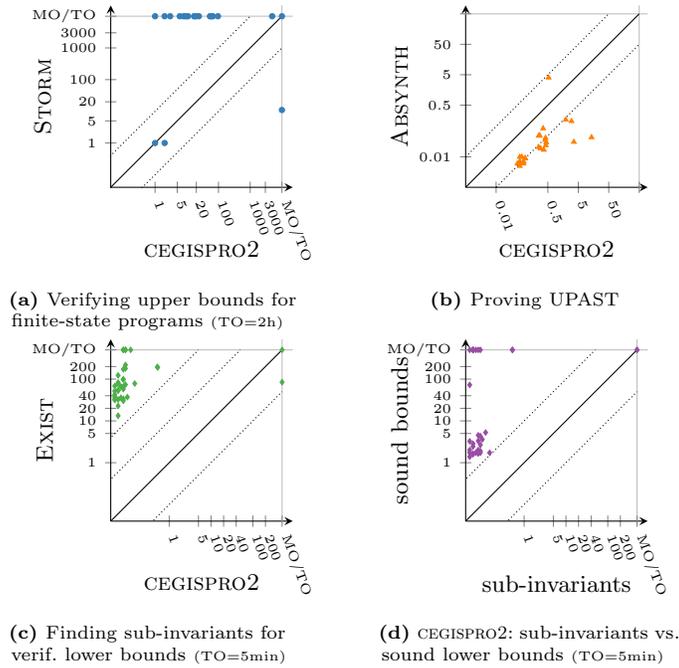

	\centering
	\setlength{\scatterplotsize}{0.328\linewidth}
	\begin{subfigure}[t]{0.328\textwidth}
		\captionsetup[subfigure]{margin=0\columnwidth}
		\centering
		\subfloat[Verifying upper bounds for \\ finite-state programs {\tiny(TO=2h)}\label{scat:storm}]{ %
			\scatterplotstorm{plotdata/scatter_storm.csv}{cegis}{\tool}{storm}{\headername{Storm}}{false}
		}
	\end{subfigure}
	\hspace*{.5cm}
	\begin{subfigure}[t]{0.328\textwidth}
		\captionsetup[subfigure]{margin=0.18\columnwidth}
		\centering
		\subfloat[Proving UPAST \label{scat:absynth}]{
			\scatterplotabsynth{plotdata/scatter_absynth.csv}{cegis}{\tool}{absynth}{\headername{Absynth}}{false}
		}
	\end{subfigure}\\
	\begin{subfigure}[t]{0.328\textwidth}
		\captionsetup[subfigure]{margin=0\columnwidth}
		\centering
		\subfloat[Finding sub-invariants for \\ verif.\ lower bounds {\tiny(TO=5min)}\label{scat:exist}]{
			\scatterplotexist{plotdata/scatter_exist_unsound.csv}{cegis}{\tool}{exist}{\headername{Exist}}{false}
		}
	\end{subfigure}
	\hspace*{.5cm}
	\begin{subfigure}[t]{0.328\textwidth}
		\captionsetup[subfigure]{margin=0.05\columnwidth}
		\centering
			\subfloat[\tool: sub-invariants vs.\ sound lower bounds {\tiny(TO=5min)}\label{scat:cegis}]{
				\scatterplotexistsoundunsound{plotdata/scatter_exist_cegis_sound_unsound.csv}{cegisunsound}{sub-invariants}{cegissound}{sound bounds}{false}
			}
	\end{subfigure}
%
	%
	\caption{Performance of \tool vs.\ state-of-the-art tools on three verification tasks (time in seconds, log-scaled; MO=8GB). Markers above the solid line depict benchmarks where \tool is faster (in different orders of magnitude marked by the \mbox{dashed lines)}.
	}
	\label{fig:scat}
\end{figure}

\medskip\noindent%
\textbf{Finite-State Programs.}
\Cref{scat:storm} depicts experiments on verifying upper bounds on expected values of finite-state programs.
    For each benchmark, \ie program and property with increasingly sharper bounds, we evaluate {\tool} on all template-refinement strategies (\cf~\Cref{sec:refinement}). 
We compare explicit- and symbolic-state engines of 
the probabilistic model checker \toolname{Storm} 1.6.3~\cite{DBLP:journals/corr/abs-2002-07080} with exact arithmetic. \toolname{Storm} implements LP-based model checking (as in \Cref{sec:oneshot}) but employs more efficient methods in its default configuration.
\Cref{scat:storm} depicts the runtime of the best configuration. See detailed configurations \mbox{in \Cref{sec:storm} (\Cref{tab:storm}).}

\smallskip\noindent%
\textit{Results.}
\begin{enumerate*}[label=(\roman*)]
	\item Our CEGIS approach synthesizes inductive invariants for a variety of programs. We mostly find \emph{syntactically small} invariants with a \emph{small number of counterexamples} compared to the state-space size (\cf~\Cref{tab:storm}). This indicates that piecewise linear inductive invariants can be sufficiently expressive for the verification of finite-state programs. The overall performance of \tool depends highly on the sharpness of the given thresholds. 
	\item Our approach can outperform state-of-the-art \emph{explicit- and symbolic-state} model checking techniques and can scale to huge state spaces. There are also simple programs where our method fails to find an inductive invariant (\expname{gridbig}) or finds inductive invariants only for rather simple properties while requiring many counterexamples (\expname{gridsmall}). Whether we need more sophisticated template refinements or whether these programs are not amenable to piecewise linear expectations is left for future work.
	\item There is no clear winner between the two fixed-partition template-refinement strategies (\cf~\Cref{tab:storm}). We further observe that the non-fixed-partition refinement is not competitive as significantly more time is spent in the synthesizer 
	to solve formulae with Boolean structures. We thus conclude that searching for good fixed-partition templates in a separate outer loop (\cf~\cref{fig:framework}) pays off.
\end{enumerate*}

\medskip\noindent%
\textbf{Proving UPAST.}
\Cref{scat:absynth} depicts experiments on proving UPAST of (possibly infinite-state) programs taken from \cite{DBLP:conf/pldi/NgoC018} (restricted to $\Nats$-valued, linear programs with flattened nested loops).
We compare to the LP-based tool \headername{Absynth} \cite{DBLP:conf/pldi/NgoC018} for computing upper bounds on expected runtimes. These benchmarks do not require template refinements. More details are given in \Cref{sec:absynth}.

\smallskip\noindent%
\textit{Results.}
\tool can prove UPAST of various infnite-state programs from the literature using very few counterexamples. \headername{Absynth} mostly outperforms \tool\footnote{\headername{Absynth} uses 
floating-point arithmetic whereas \tool 
uses exact arithmetic.}, which is to be expected as \headername{Absynth} is tailored to the computation of expected runtimes. Remarkably, the runtime bounds synthesized by \tool are often as tight as the bounds synthesized by \headername{Absynth} (\cf~\Cref{tab:absynth}).
 
\medskip\noindent%
\textbf{Verifying Lower Bounds.}
\Cref{scat:exist} depicts experiments aiming to verify lower bounds on expected values of (possibly infinite-state) programs taken from \cite{DBLP:conf/cav/BaoTPHR22}. We compare to \headername{Exist} \cite{DBLP:conf/cav/BaoTPHR22}\footnote{\headername{Exist} supports parametric probabilities, which are not supported by our tool. We have instantiated these parameters with varying probabilities to enable a comparison.}, which combines CEGIS with sampling- and ML-based techniques. However, \headername{Exist} synthesizes sub-invariants only, which might be unsound for proving lower bounds (cf.\ \Cref{sec:past_lower}). Thus, for a fair comparison, \Cref{scat:exist} depicts experiments where \emph{both} \headername{Exist} and \tool synthesize sub-invariants only, whereas in \Cref{scat:cegis}, we compare \tool that finds sub-invariants only with \tool that \emph{additionally} proves UPAST and c.d.b., thus obtaining sound lower bounds as per \Cref{thm:lowerbounds}. 
No benchmark requires template refinements.

\smallskip\noindent%
\textit{Results.}
\tool is capable of verifying quantitative lower bounds and outperforms \headername{Exist} (on 30/32 benchmarks) for synthesizing sub-invariants. Additionally proving UPAST and c.d.b.\ naturally requires more time. A manual inspection reveals that, for most TO/MO cases in \cref{scat:cegis}, there is no c.d.b.\ sub-invariant. 
One soundness check times out, since we could not prove UPAST for that benchmark.

	\section{Related Work}
\label{sec:related}

We discuss related works in invariant synthesis, probabilistic model checking, and symbolic inference.
ICE~\cite{ICE} is a template-based, cex.-guided technique for learning invariants. More inductive synthesis approaches are surveyed in~\cite{DBLP:series/natosec/AlurBDF0JKMMRSSSSTU15,DBLP:conf/tacas/FedyukovichB18}.

\smallskip\noindent%
\textit{Quantitative Invariant Synthesis.}
%
%
Apart from the discussed method \cite{DBLP:conf/cav/BaoTPHR22}, \emph{constraint solving-based approaches} \cite{DBLP:conf/atva/FengZJZX17,DBLP:conf/cav/ChenHWZ15,DBLP:conf/sas/KatoenMMM10} aim to synthesize quantitative invariants for proving lower bounds over $\Reals$-valued program variables -- arguably a simplification as it allows solvers to use (decidable) real arithmetic. 
In particular,~\cite{DBLP:conf/cav/ChenHWZ15} also obtains linear constraints from counterexamples ensuring certain validity conditions on candidate invariants. Apart from various technical differences, we identify three conceptual differences:
\begin{enumerate*}[label=(\roman*)]
	\item we support piecewise expectations which have been shown sufficiently expressive for verifying quantitative reachability properties; 
	\item we focus on the integration of fast verifiers over efficiently decidable theories; and 
	\item we do not need to \emph{assume} termination or \mbox{boundedness of expectations.}
\end{enumerate*}
%
 

Various \emph{martingale-based approaches}, such as \cite{DBLP:conf/cav/ChakarovS13,DBLP:conf/cav/ChatterjeeFG16,DBLP:conf/popl/ChatterjeeNZ17,DBLP:conf/vmcai/FuC19,DBLP:conf/popl/FioritiH15,DBLP:journals/pacmpl/AgrawalC018,DBLP:conf/esop/MoosbruggerBKK21}, aim to synthesize quantitative invariants over $\Reals$-valued variables, see \cite{DBLP:journals/toplas/TakisakaOUH21} for a recent survey. 
Most of these approaches yield invariants for proving almost-sure termination or bounding expected runtimes.
\emph{$\varepsilon$-decreasing supermartingales}~\cite{DBLP:conf/cav/ChakarovS13,DBLP:conf/tacas/ChakarovVS16} and \emph{nonnegative repulsing supermartingales}~\cite{DBLP:journals/toplas/TakisakaOUH21} can upper-bound arbitrary reachability probabilities.
In contrast, we synthesize invariants for proving upper- or lower bounds for more general quantities, \ie expectations. 
\cite{DBLP:conf/cav/BartheEFH16} can prove bounds on expected values via symbolic reasoning and \emph{Doob's decomposition}, which, however, requires user-supplied invariants and hints. 
\cite{DBLP:conf/cav/AbateGR20} employs a CEGIS loop to train a neural network dedicated to learning a ranking supermartingale witnessing UPAST of (possibly continuous) probabilistic programs. They also use counterexamples provided by SMT solvers to guide the learning process.

The \emph{recurrence solving-based approach} in~\cite{DBLP:conf/atva/BartocciKS19} synthesizes nonlinear invariants encoding 
(higher-order) moments of program variables.
However, the underlying algebraic techniques are confined to the sub-class of \emph{prob-solvable loops}. 

\smallskip\noindent%
\textit{Probabilistic Model Checking.}
Symbolic probabilistic model checking focusses mostly on algebraic decision diagrams~\cite{DBLP:conf/icalp/BaierCHKR97,DBLP:conf/tacas/AlfaroKNPS00}, representing the transition relation symbolically and using equation solving or value iteration \cite{DBLP:conf/cav/Baier0L0W17,DBLP:conf/cav/HartmannsK20,DBLP:conf/cav/QuatmannK18} on that representation. 
PrIC3~\cite{pric3} 
finds quantitative invariants by iteratively overapproximating $k$-step reachability.
Alternative CEGIS approaches synthesize Markov chains~\cite{DBLP:journals/fac/0002HJK21} and probabilistic programs~\cite{DBLP:conf/tacas/Andriushchenko021} that satisfy reachability properties.

\smallskip\noindent%
\textit{Symbolic Inference.}
Probabilistic inference -- in the finite-horizon case -- employs weighted model counting via either decision diagrams annotated with probabilities as in \toolname{Dice}~\cite{DBLP:journals/pacmpl/HoltzenBM20,DBLP:conf/cav/HoltzenJVMSB20} or approximate versions by SAT/SMT-solvers~\cite{DBLP:conf/ijcai/ChakrabortyFMV15,DBLP:conf/aaai/ChakrabortyMMV16,DBLP:journals/acta/ChistikovDM17,DBLP:conf/qest/RabeWKYH14,DBLP:conf/ijcai/BellePB15}. 
\toolname{PSI}~\cite{DBLP:conf/cav/GehrMV16} determines symbolic representations of exact distributions.
\toolname{Prodigy}~\cite{DBLP:conf/cav/ChenKKW22} decides whether a 
probabilistic loop agrees with an (invariant) specification.

	
	\begin{acknowledgement}
		The datasets generated during and/or analysed during the current study are available in the Zenodo repository~\cite{batz_kevin_2023_7507921}.
	\end{acknowledgement}
	
	\bibliographystyle{splncs04}
	\bibliography{literature}
	
	\allowdisplaybreaks 
	\appendix

\section{Symbolically Constructing Characteristic Functions}
\label{app:wp}

Fix a probabilistic loop $\WHILEDO{\Guard}{\Body}$
and a postexpectation $f \in \E$.
We employ a \emph{weakest preexpectation} à la McIver and Morgan~\cite{McIverM05} to symbolically compute the expectation $\wp{\Body}{f} \in \E$ that maps every state $\state \in \States$ to the expected value of postexpectation $f$ evaluated in the final states reached after executing loop body $\Body$ when started in state $\state$.
Formally, the weakest preexpectation transformer $\wpsymbol\colon \Body \to (\E \to \E)$ is given by the rules in \Cref{table:wp}. 
The characteristic function $\charfun{f}$ of a probabilistic loop 
$\WHILEDO{\guard}{C}$ with respect to postexpectation $f$ is %
\begin{align*}
	\charfun{f}(X) \eeq \iverson{\neg \guard} \cdot f \pplus \iverson{\guard}\cdot \wp{C}{X}~.
\end{align*}
The symbolic template transformer $\tcharfun{f}$ from \Cref{thm:templatetrans} is defined analogously for templates instead of arbitrary expectations (notice that all transformations of $\wpsymbol$ are syntactic).
Every piecewise linear expectation in $\Elin$ (or template in $\Temp$) can then be converted into a template using the transformation into guarded normal form in \cite{DBLP:conf/cav/BatzCKKMS20}, where template variables are treated as constants.

\begin{table}[h]
	
	\renewcommand{\arraystretch}{1.5}
	
	\begin{center}
		\begin{tabular}{@{\hspace{1em}}l@{\hspace{2em}}l}
			\toprule
			$\boldsymbol{\cc}$			& $\boldsymbol{\textbf{\textsf{wp}}\,\left \llbracket \cc\right\rrbracket  \left(f \right)}$ \\
			\midrule
			$\SKIP$					& $f$ 																					\\
			$\ASSIGN{x}{\ee}$			& $f\subst{x}{\ee}$ \\
			$\COMPOSE{\cc_1}{\cc_2}$		& $\wp{\cc_1}{\vphantom{\big(}\wp{\cc_2}{f}}$ \\
			$\PCHOICE{\cc_1}{\pp}{\cc_2}$		& $\pp \cdot \wp{\cc_1}{f} + (1- \pp) \cdot \wp{\cc_2}{f}$ \\
			$\ITE{\guard}{\cc_1}{\cc_2}$		& $\iverson{\guard} \cdot \wp{\cc_1}{f} + \iverson{\neg \guard} \cdot \wp{\cc_2}{f}$ \\[.25em]
			%
			\bottomrule
		\end{tabular}
	\end{center}
	\caption{Rules defining the weakest preexpectation transformer for loop-free programs. 
		$\iverson{\varphi}$ denotes an indicator function, \ie $\iverson{\varphi}(s) = 1$ if $s \models \varphi$ and $0$ otherwise. Moreover, $f\subst{x}{\ee} = \lambda \state\mydot f(s[x \mapsto \ee(s)])$ denotes the expectation $f$ in which every free occurrences of variable $x$ are substituted by expression $\ee$.}
	\label{table:wp}
\end{table}

Notice that, as long as only finitely many states are reachable, one can alternatively construct the characteristic function by first building the Markov chain underlying a probabilistic loop~\cite{DBLP:journals/pe/GretzKM14}, and then deriving the well-known Bellman operator (\cf \cite{Puterman1994Markov}). 
Furthermore, if $f$ is a predicate (\ie maps states to $0$ or $1$), the characteristic function corresponds to the least fixed point characterization of reachability probabilities~\cite[Thm.~10.15]{Baier2008Principles}.%



\section{Proof of \Cref{thm:satforindinv}}
\subsubsection*{Claim.}
	For every 
	template $\temp \in \Ntemp$ and $f \in \Elin$, we have
	\begin{align*}
&\generator{\temp} \cap \indinvshort \neq \emptyset\\
	\textnormal{iff} \quad &
	 \satset\big(\:\forall\state \in \States_\Guard \colon \quad
	\underbrace{0 \leq \temp(\state)}_{\mathclap{\text{well-definedness}}} \wwedge \underbrace{\tcharfun{f}(\temp)(\state) \leq \temp(\state)}_{\text{inductivity}} \wwedge \underbrace{\temp(\state) \leq g(\state)}_{\text{safety}}\:\big)
	~\neq~ \emptyset. 
	\end{align*}
\begin{proof}
Recall the definition of $\indinvshort$ from our problem statement.
Then:
\begin{align*}
	&\generator{\temp} \cap \indinvshort \neq \emptyset\\
	\text{iff} \quad &\exists \inv \in \generator{\temp}\colon \inv \in \indinvshort \\
    \text{iff} \quad &\exists \inv \in \generator{\temp}\colon
    	0 \preceq \inv \wedge \charfun{f}(\inv) \preceq \inv \wedge \inv \preceq g  
    	\tag{see formal problem statement}\\
     \text{iff} \quad &\exists \inv \in \generator{\temp}	\colon \forall\state \in \States \colon 
    0 \leq \inv(\state) \wedge \charfun{f}(\inv)(\state) \leq \inv(\state) \wedge \inv(\state) \leq g(\state) 
    \\
    \text{iff} \quad &\exists \tval 	\colon \forall \state \in \States \colon 
    0 \leq \tvalapps{\temp}(\state) \wedge \charfun{f}(\tvalapps{\temp})(\state) \leq \tvalapps{\temp}(\state) \wedge \tvalapps{\temp}(\state) \leq g(\state)  \\
    \text{iff} \quad &\exists \tval 	\colon \forall \state \in \States_\Guard \colon 
    0 \leq \tvalapps{\temp}(\state) \wedge \charfun{f}(\tvalapps{\temp})(\state) \leq \tvalapps{\temp}(\state) \wedge \tvalapps{\temp}(\state) \leq g(\state)  
    \tag{since $T$ is natural and $\lfp \charfun{f} \preceq g$ by assumption}\\
    \text{iff} \quad &\exists \tval 	\colon \forall\state \in \States_\Guard  \colon 
    0 \leq \tvalapps{\temp}(\state) \wedge \tvalapps{\tcharfun{f}(\temp)}(\state) \leq \tvalapps{\temp}(\state) \wedge\tvalapps{\temp}(\state) \leq g(\state)  
    \tag{by \Cref{thm:templatetrans}}\\
    \text{iff} \quad &\satset\left(\forall\state\in \States_\Guard \colon 
    0 \leq \temp(\state) \wedge \tcharfun{f}(\temp)(\state) \leq \temp(\state) \wedge \temp(\state) \leq g(\state)\right)\neq \emptyset~.
    \tag*{\qed}
    \end{align*}
\end{proof}%

%

\section{Cooperative Verifiers}\label{app:cooperative}
The baseline verifier from \Cref{def:verifier} may return any counterexample. Without further restrictions, a verifier may start enumerating these counterexamples, which may yield a bad performance. \mbox{For instance, let $f = \iverson{c=1}$,}
 \begin{align*}
 	&\WHILEDO{c=0 \wedge x<100}{\PCHOICE{\ASSIGN{x}{x+1}}{0.99}{\ASSIGN{c}{1}}},~\text{and} \\
 	&\temp \eeq  \iverson{c=0 \wedge x<100}\cdot (\alpha \cdot c + \beta \cdot x + \gamma ) + \iverson{c=1} + \iverson{c=0 \wedge x\geq 100}\cdot 0~.
 \end{align*}%
Assume that the verifier returns two consecutive states $s_1,s_2$ with $s_i(c)= 0$ and $s_i(x) =i$ for $i \in\{1,2\}$. The constraints ensuring inductivity at $s_1$ and $s_2$ are%
\begin{align*}
	\textcolor{gray}{\tcharfun{f}(\temp)(s_i) \quad\! =} \quad  0.99\cdot((i+1)\cdot\beta + \gamma) + 0.01 \lleq i\cdot \beta + \gamma \quad \textcolor{gray}{= \quad T(s_i)}~.
	%
	%
\end{align*}%
The two constraints are very similar and there is only little information gain from the constraint obtained from $s_2$. Constraints obtained from more diverse states such as $\state'$ with, \eg $\state'(x)=98$, prune many more undesired template instances.
%
%
\noindent%

We are therefore interested in cooperative verifiers -- \emph{teachers} -- that return sufficiently diverse counterexamples. 
We tackle diversity by defining a distance measure $\mu\colon \States \times \States \rightarrow \Reals$ on states\footnote{In our implementation, we use the Manhattan distance on the program variables.} and provide the verifier \emph{additionally} to $I$ with the last counterexample $s'$ and a lower bound $m$.
The cooperative verifier then preferably returns a new counterexample $\state$ such that $\mu(\state, \state') \geq m$.%
\begin{definition}
\label{def:cverifier}
A \emph{cooperative verifier}  
\mbox{$\coopverify_\mu \colon \States \times \Reals \to \Elin \to \{\true\} \cup \States$} is a function such that
(i) $\coopverify_\mu(\state_\text{last}, m)$ is a verifier for all $\state_\text{last} \in \States$ and $m \in \Reals$, 
	%
and (ii) $\coopverify_\mu(\state_\text{last}, m)(I) = \state$ implies either (a) $\mu(\state,\state_\text{last}) \geq m$ or (b) for all $\hat{\state} \in \counterex{\inv}$, $\mu(\state_\text{last},\hat{\state}) \leq m$.
\end{definition}%
Since every cooperative verifier refines the contract of a verifier, our correctness and completenes results carry over.
%
Furthermore, we can easily generalize the cooperative verifier beyond taking into account only the last counterexample.
In our implementation, rather than a fixed value $m$, we adapt $m$ during runtime: If we succeed in finding two counterexamples that were $m$ apart, we update $m \gets m \cdot d$, otherwise,  $m \gets m \cdot \nicefrac{1}{d}$ for suitable values of $d$.

\section{Additional Details on Template Refinement}


\label{sec:refinement}
Our problem statement requires a \emph{given} template $\temp$ and selecting templates is hard.
If $T$ is too restrictive, it excludes admissible invariants, \ie \mbox{$\generator{\temp}\cap\indinvshort = \emptyset$}. 
If $T$ is too liberal, the synthesizer has to search a high-dimensional space. 

We take an optimistic approach as depicted in \Cref{fig:framework}:
%
we start with a small restrictive template $\temp_1$. 
If we learn that $\temp_1$ contains no admissible invariant, we ask the template generator for a new template $T_2$, and so on, until we eventually (and hopefully) find a small template $\temp_{i}$ with $\generator{\temp_i}\cap\indinvshort \neq \emptyset$. 
Below, we investigate three heuristics. 
While the approaches do \emph{not} yield step-wise refinements, \ie $\generator{T_i} \subseteq \generator{T_{i+1}}$, all approaches ensure \emph{progress}, \ie $\generator{T_i} \subsetneq \generator{T_{i+j}}$ for some $j \geq 1$. 
For finite-state programs, progress ensures completeness: we eventually reach a maximally-partitioned template $T$ in which every state has its own piece; $\generator{\temp}$ then contains at least one admissible invariant, namely $\lfp \charfun{f}$.


\smallskip\noindent%
\textit{Static Hyperrectangle Refinement for Finite-State Programs.} 
For \emph{finite-state} programs, we obtain lower- and upper bounds on each variable. Hence, the state space is (a subset of) a bounded hyperrectangle. We obtain template $\temp_i$ for $i>1$ by splitting each dimension of this hyperrectangle into $i$ equally-sized parts\footnote{if possible, \ie if $i$ does not exceed the size of the dimension.}, thus obtaining (at most) $i^{\sizeof{\Vars}}$ new hyperrectangles. 
Let the hyperrectangles be described by expressions $R_1,\hdots,R_m$,
and assume that the initial template is $\temp_1 = \sum_{j=1}^{n} \iverson{\Tb_j}\cdot \Te_j$. Then, we obtain $\temp_i$ as $\temp_i = \sum_{j=1}^n \sum^m_k \iverson{\Tb_j \land R_k} \cdot \Te_{k,j}$.

\smallskip\noindent%
\textit{Dynamic Hyperrectangle Refinement.} 
We proceed as for static refinement but do 
not fix \emph{where} we split the 
hyperrectangle, \ie given a possibly unbounded hyperrectangle capturing the (now possibly infinite)
state space, we introduce template variables encoding the boundaries of the hyperrectangles, as in \Cref{eq:varpartemplexample}.
%

\smallskip\noindent%
\textit{Inductivity-Guided Refinement.} 
We refine templates using a hint, particularly the last partially admissible instance $\inv$ in the synthesizer concerning a fixed-partition template $\temp = \sum_{j=1}^{n} \iverson{\Tb_j}\cdot \Te_j$. We split every $\Tb_j$ into those parts where $\inv$ is partially inductive, \ie $\tcharfun{f}(\inv) \preceq \inv$, and where it is not.
This partitioning can be computed symbolically using the construction from~\cite[Thm.~7]{DBLP:conf/cav/BatzCKKMS20}. 
%
%
%

\section{Additional Details on Experiments}
\label{app:empricial}

The finite-state (\Cref{scat:storm}) benchmarks were conducted on a 2.1GHz Intel Xeon processor (one core per benchmark). All other benchmarks were conducted on a 2.3 GHz Dual-Core Intel Core i5.


\subsection{Details on the Comparison of \tool and \toolname{Storm}}
\label{sec:storm}

\noindent
Table \ref{tab:storm} depicts the results for finite-state programs; the corresponding programs are depicted in \Cref{sec:programs_storm}.  We employ a \emph{cooperative verifier} with $d=2$. Column \headername{Prog} depicts the name of the program and $S$ is the program's state-space size. \headername{sp} and \headername{dd} are the runtimes of \toolname{Storm}'s sparse and decision diagram-based engines, respectively. \headername{best} is the best runtime under all CEGIS configurations, which are then listed. We show the number $|S'|$ of counterexamples, the size $|I|$ of the inductive invariant (in terms of the number of linear piecewise), the fraction $t_\text{s}${\small{\%}} of time spent in the synthesizer, and the total time $t$.

\begin{table}[h]
	\centering
	\caption{\toolname{Storm} vs.\ \tool on finite-state programs (TO=2h, MO=8GB. Time in seconds).}\label{tab:storm}
	\adjustbox{width=0.8\textwidth}{

\begin{tabular}{lc||rr||r|rrrr|rrrr|rrrr}
	&   \multicolumn{1}{c}{}& \multicolumn{2}{c}{\headername{Storm}} & \multicolumn{1}{c}{} & \multicolumn{12}{c}{\headername{\tool}}  \\ 
	&   & && & \multicolumn{4}{c|}{\headername{induct.-guided}} & \multicolumn{4}{c|}{\headername{static}} & \multicolumn{4}{c}{\headername{dynamic}}  \\ 
	Prog & $|S|$   & {\headername{sp}} & {\headername{dd}} & {\headername{best}} & {\footnotesize $|S'|$} &{\footnotesize $|\inv|$} &  {\footnotesize $\text{t}_\text{s}$\%} & {\footnotesize t} & {\footnotesize $|S'|$} & {\footnotesize $|\inv|$ } & {\footnotesize $\text{t}_\text{s}$\%}  & {\footnotesize t } & {\footnotesize $|S'|$} & {\footnotesize $|\inv|$ } & {\footnotesize $\text{t}_\text{s}$\%}  & {\footnotesize t  } \\ 
	\midrule\midrule\multirow{3}{*}{\footnotesize{boundedrwmultistep}} & \multirow{3}{*}{{\scriptsize $1\cdot10^5$}}  & {\scriptsize MO}&{\scriptsize TO}&{\scriptsize {\scriptsize $3$}}&{\scriptsize $33$}&{\scriptsize $10$}&{\scriptsize $40$}&{\scriptsize $3$}&--&--&--&{\scriptsize TO}&--&--&--&{\scriptsize TO}\\ 
	&  & {\scriptsize MO}&{\scriptsize TO}&{\scriptsize {\scriptsize $10$}}&{\scriptsize $55$}&{\scriptsize $16$}&{\scriptsize $36$}&{\scriptsize $10$}&--&--&--&{\scriptsize TO}&--&--&--&{\scriptsize TO}\\ 
	&  & {\scriptsize MO}&{\scriptsize TO}&--&--&--&--&{\scriptsize TO}&--&--&--&{\scriptsize TO}&--&--&--&{\scriptsize TO}\\ 
	\midrule  
	\multirow{3}{*}{\footnotesize{brp}} & \multirow{3}{*}{{\scriptsize $1\cdot10^{10}$}} & {\scriptsize MO}&{\scriptsize TO}&{\scriptsize {\scriptsize $11$}}&{\scriptsize $56$}&{\scriptsize $23$}&{\scriptsize $40$}&{\scriptsize $11$}&{\scriptsize $70$}&{\scriptsize $10$}&{\scriptsize $35$}&{\scriptsize $18$}&--&--&--&{\scriptsize TO}\\ 
	&  & {\scriptsize MO}&{\scriptsize TO}&{\scriptsize {\scriptsize $54$}}&{\scriptsize $138$}&{\scriptsize $42$}&{\scriptsize $63$}&{\scriptsize $253$}&{\scriptsize $125$}&{\scriptsize $17$}&{\scriptsize $27$}&{\scriptsize $54$}&--&--&--&{\scriptsize TO}\\ 
	&  & {\scriptsize MO}&{\scriptsize TO}&{\scriptsize {\scriptsize $56$}}&{\scriptsize $104$}&{\scriptsize $41$}&{\scriptsize $54$}&{\scriptsize $111$}&{\scriptsize $122$}&{\scriptsize $17$}&{\scriptsize $30$}&{\scriptsize $56$}&--&--&--&{\scriptsize TO}\\ 
	\midrule  
	\multirow{3}{*}{\footnotesize{brpfinitefamily}} &  \multirow{3}{*}{{\scriptsize $16\cdot10^{13}$}}& {\scriptsize TO}&{\scriptsize TO}&{\scriptsize {\scriptsize $8$}}&{\scriptsize $53$}&{\scriptsize $7$}&{\scriptsize $72$}&{\scriptsize $10$}&{\scriptsize $67$}&{\scriptsize $7$}&{\scriptsize $44$}&{\scriptsize $8$}&{\scriptsize $54$}&{\scriptsize $9$}&{\scriptsize $52$}&{\scriptsize $29$}\\ 
	&  & {\scriptsize TO}&{\scriptsize TO}&{\scriptsize {\scriptsize $17$}}&{\scriptsize $64$}&{\scriptsize $13$}&{\scriptsize $74$}&{\scriptsize $17$}&{\scriptsize $215$}&{\scriptsize $19$}&{\scriptsize $66$}&{\scriptsize $373$}&--&--&--&{\scriptsize TO}\\ 
	&  & {\scriptsize TO}&{\scriptsize TO}&{\scriptsize {\scriptsize $18$}}&{\scriptsize $68$}&{\scriptsize $12$}&{\scriptsize $68$}&{\scriptsize $18$}&{\scriptsize $231$}&{\scriptsize $19$}&{\scriptsize $80$}&{\scriptsize $731$}&--&--&--&{\scriptsize TO}\\ 
	\midrule  
	\multirow{3}{*}{\footnotesize{chain}} &\multirow{3}{*}{{\scriptsize $1\cdot10^{12}$}}  & {\scriptsize MO}&{\scriptsize TO}&{\scriptsize {\scriptsize $1$}}&{\scriptsize $97$}&{\scriptsize $6$}&{\scriptsize $68$}&{\scriptsize $10$}&{\scriptsize $66$}&{\scriptsize $3$}&{\scriptsize $50$}&{\scriptsize $1$}&--&--&--&{\scriptsize TO}\\ 
	&  & {\scriptsize MO}&{\scriptsize TO}&{\scriptsize {\scriptsize $24$}}&--&--&--&{\scriptsize TO}&{\scriptsize $116$}&{\scriptsize $5$}&{\scriptsize $86$}&{\scriptsize $24$}&--&--&--&{\scriptsize TO}\\ 
	&  & {\scriptsize MO}&{\scriptsize TO}&{\scriptsize {\scriptsize $4933$}}&--&--&--&{\scriptsize TO}&{\scriptsize $503$}&{\scriptsize $23$}&{\scriptsize $81$}&{\scriptsize $4933$}&--&--&--&{\scriptsize TO}\\ 
	\midrule  
	\multirow{3}{*}{\footnotesize{chainselectstepsize}} &\multirow{3}{*}{{\scriptsize $3\cdot10^7$}}  & {\scriptsize MO}&{\scriptsize TO}&{\scriptsize {\scriptsize $9$}}&{\scriptsize $156$}&{\scriptsize $7$}&{\scriptsize $71$}&{\scriptsize $29$}&{\scriptsize $156$}&{\scriptsize $7$}&{\scriptsize $71$}&{\scriptsize $29$}&{\scriptsize $81$}&{\scriptsize $7$}&{\scriptsize $49$}&{\scriptsize $9$}\\ 
	&  & {\scriptsize MO}&{\scriptsize TO}&{\scriptsize {\scriptsize $96$}}&--&--&--&{\scriptsize TO}&{\scriptsize $179$}&{\scriptsize $15$}&{\scriptsize $70$}&{\scriptsize $96$}&--&--&--&{\scriptsize TO}\\ 
	&  & {\scriptsize MO}&{\scriptsize TO}&{\scriptsize {\scriptsize $66$}}&--&--&--&{\scriptsize TO}&{\scriptsize $164$}&{\scriptsize $15$}&{\scriptsize $58$}&{\scriptsize $66$}&--&--&--&{\scriptsize TO}\\ 
	\midrule  
	\multirow{1}{*}{\footnotesize{gridbig}} &\multirow{1}{*}{{\scriptsize $1\cdot10^6$}}  & {\scriptsize {\scriptsize $11$}}&{\scriptsize --}&--&--&--&--&{\scriptsize TO}&--&--&--&{\scriptsize TO}&--&--&--&{\scriptsize TO}\\ 
	\midrule  
	\multirow{2}{*}{\footnotesize{gridsmall}} & \multirow{2}{*}{{\scriptsize $1\cdot10^2$}} & {\scriptsize {\scriptsize ${<}1$}}&{\scriptsize {\scriptsize $32$}}&{\scriptsize {\scriptsize $1$}}&{\scriptsize $15$}&{\scriptsize $7$}&{\scriptsize $36$}&{\scriptsize $1$}&{\scriptsize $46$}&{\scriptsize $10$}&{\scriptsize $37$}&{\scriptsize $3$}&{\scriptsize $20$}&{\scriptsize $5$}&{\scriptsize $39$}&{\scriptsize $2$}\\ 
	&  & {\scriptsize {\scriptsize ${<}1$}}&{\scriptsize {\scriptsize $32$}}&{\scriptsize {\scriptsize $2$}}&{\scriptsize $26$}&{\scriptsize $11$}&{\scriptsize $35$}&{\scriptsize $2$}&{\scriptsize $77$}&{\scriptsize $17$}&{\scriptsize $32$}&{\scriptsize $10$}&{\scriptsize $71$}&{\scriptsize $10$}&{\scriptsize $82$}&{\scriptsize $59$}\\ 
	\midrule  
	\multirow{3}{*}{\footnotesize{zeroconf}} & \multirow{3}{*}{{\scriptsize $1\cdot10^8$}}  & {\scriptsize MO}&{\scriptsize TO}&{\scriptsize {\scriptsize ${<}1$}}&{\scriptsize $7$}&{\scriptsize $3$}&{\scriptsize $22$}&{\scriptsize ${<}1$}&{\scriptsize $7$}&{\scriptsize $3$}&{\scriptsize $26$}&{\scriptsize ${<}1$}&{\scriptsize $7$}&{\scriptsize $3$}&{\scriptsize $31$}&{\scriptsize ${<}1$}\\ 
	&  & {\scriptsize MO}&{\scriptsize TO}&{\scriptsize {\scriptsize ${<}1$}}&{\scriptsize $105$}&{\scriptsize $22$}&{\scriptsize $60$}&{\scriptsize $32$}&{\scriptsize $9$}&{\scriptsize $5$}&{\scriptsize $39$}&{\scriptsize ${<}1$}&{\scriptsize $156$}&{\scriptsize $7$}&{\scriptsize $92$}&{\scriptsize $215$}\\ 
	&  & {\scriptsize MO}&{\scriptsize TO}&{\scriptsize {\scriptsize $63$}}&--&--&--&{\scriptsize TO}&{\scriptsize $173$}&{\scriptsize $23$}&{\scriptsize $59$}&{\scriptsize $63$}&--&--&--&{\scriptsize TO}\\ 
	\midrule  
	\multirow{3}{*}{\footnotesize{zeroconffamily}} & \multirow{3}{*}{{\scriptsize $1\cdot10^{16}$}}  & {\scriptsize TO}&{\scriptsize TO}&{\scriptsize {\scriptsize $2$}}&{\scriptsize $48$}&{\scriptsize $3$}&{\scriptsize $49$}&{\scriptsize $2$}&{\scriptsize $48$}&{\scriptsize $3$}&{\scriptsize $49$}&{\scriptsize $2$}&{\scriptsize $59$}&{\scriptsize $3$}&{\scriptsize $52$}&{\scriptsize $3$}\\ 
	&  & {\scriptsize TO}&{\scriptsize TO}&{\scriptsize {\scriptsize $6$}}&--&--&--&{\scriptsize TO}&{\scriptsize $77$}&{\scriptsize $9$}&{\scriptsize $56$}&{\scriptsize $6$}&{\scriptsize $252$}&{\scriptsize $13$}&{\scriptsize $81$}&{\scriptsize $854$}\\ 
	&  & {\scriptsize TO}&{\scriptsize TO}&{\scriptsize {\scriptsize $20$}}&--&--&--&{\scriptsize TO}&{\scriptsize $99$}&{\scriptsize $9$}&{\scriptsize $70$}&{\scriptsize $20$}&{\scriptsize $164$}&{\scriptsize $13$}&{\scriptsize $73$}&{\scriptsize $265$}\\ 
	\midrule  
\end{tabular}
}
\end{table}

\FloatBarrier

\subsection{Details on the Comparison of Absynth and \tool}
\label{sec:absynth}

\Cref{tab:absynth} depicts the results on the UPAST benchmarks. \headername{Prog} is the name of the benchmark, t denotes the runtime in seconds required by the tools and \headername{bound} indicates their respective synthesized bounds. $|S'|$ depicts the number of counterexamples required by \tool.
\begin{table}[h]
	\centering
	\caption{\toolname{Absynth} vs.\ \tool on expected runtimes (TO=20min, MO=8GB. Time in seconds).}\label{tab:absynth}
	\adjustbox{width=1.1\textwidth}{
\begin{tabular}{l||rc||rrc}
	\multicolumn{1}{c}{} &  \multicolumn{2}{c}{\headername{Absynth}} & \multicolumn{3}{c}{\headername{\tool}}  \\ 
	\multicolumn{1}{c}{\headername{Prog}}&  \multicolumn{1}{c}{{\footnotesize t}} & \multicolumn{1}{c}{{\footnotesize bound}} & \multicolumn{1}{c}{{\footnotesize $|S'|$}} & \multicolumn{1}{c}{{\footnotesize t}} & \multicolumn{1}{c}{{\footnotesize bound}} \\ 
	\midrule \midrule bayesiannetwork & {\scriptsize $0.15$} & {\scriptsize 1 + 2 max(0, n)} & {\scriptsize $1$} & {\scriptsize $2.99$} & {\scriptsize (n * 2.0)}\\ 
	\midrule  
	ber & {\scriptsize ${\leq}0.01$} & {\scriptsize 1 + 2 max(0, n - x)} & {\scriptsize $3$} & {\scriptsize $0.08$} & {\scriptsize ((x * -2.0) + (n * 3.0))}\\ 
	\midrule  
	cowboyduel & {\scriptsize ${\leq}0.01$} & {\scriptsize 1 + 1.2 max(0, flag)} & {\scriptsize $1$} & {\scriptsize $0.06$} & {\scriptsize 11/5}\\ 
	\midrule  
	C4Bt09 & {\scriptsize $0.02$} & {\scriptsize 1 + max(0, -j + x)} & {\scriptsize $15$} & {\scriptsize $0.37$} & {\scriptsize max[ ((j * -1.0) + x + 1.0), ((j * -1.0) + x + 1.0) ]}\\ 
	\midrule  
	C4Bt13 & {\scriptsize $0.02$} & {\scriptsize 1 + 2 max(0, x) + max(0, y)} & {\scriptsize $8$} & {\scriptsize $0.25$} & {\scriptsize ((x * 2.0) + y)}\\ 
	\midrule  
	C4Bt19 & {\scriptsize $0.04$} & {\scriptsize 3 + max(0, 51 + i + k) + 2 max(0, i)} & {\scriptsize $16$} & {\scriptsize $0.42$} & {\scriptsize ((i * 2.0) + k + -46.0)}\\ 
	\midrule  
	C4Bt61 & {\scriptsize $0.02$} & {\scriptsize 2 + max(0, l)} & {\scriptsize $15$} & {\scriptsize $0.43$} & {\scriptsize (l + -3.0)}\\ 
	\midrule  
	condand & {\scriptsize ${\leq}0.01$} & {\scriptsize 1 + 2 max(0, m)} & {\scriptsize $4$} & {\scriptsize $0.09$} & {\scriptsize ((m * 2.0) + 1.0)}\\ 
	\midrule  
	coupon & {\scriptsize $0.05$} & {\scriptsize 10 max(0, 5 - i)} & {\scriptsize $5$} & {\scriptsize $0.25$} & {\scriptsize max[ 149/12, 137/12, 61/6, 17/2, (i * 3/2) ]}\\ 
	\midrule  
	fcall & {\scriptsize ${\leq}0.01$} & {\scriptsize 1 + 4 max(0, n - x)} & {\scriptsize $3$} & {\scriptsize $0.09$} & {\scriptsize ((x * -3.0) + (n * 3.0))}\\ 
	\midrule  
	fillingvol & {\scriptsize $0.09$} & {\scriptsize 1 + 0.666667 max(0, 10 - vM + vTF)} & {\scriptsize $10$} & {\scriptsize $0.35$} & {\scriptsize \makecell[cc]{max[ ((vM * -1/36) + (vTF * 1/36) + 2.0),\\ ((vM * -2/3) + (vTF * 2/3) + -583/180) ]}}\\ 
	\midrule  
	geo & {\scriptsize ${\leq}0.01$} & {\scriptsize 3} & {\scriptsize $1$} & {\scriptsize $0.06$} & {\scriptsize 3.0}\\ 
	\midrule  
	linear01 & {\scriptsize ${\leq}0.01$} & {\scriptsize 1 + 0.6 max(0, x)} & {\scriptsize $1$} & {\scriptsize $0.06$} & {\scriptsize x}\\ 
	\midrule  
	trappedminer & {\scriptsize $0.03$} & {\scriptsize 1 + 11.5 max(0, -i + n)} & {\scriptsize $72$} & {\scriptsize $3.58$} & {\scriptsize ((i * -3.0) + (n * 3.0) + 1.0)}\\ 
	\midrule  
	noloop & {\scriptsize ${\leq}0.01$} & {\scriptsize 2} & {\scriptsize $1$} & {\scriptsize $0.06$} & {\scriptsize 2.0}\\ 
	\midrule  
	prdwalk & {\scriptsize $0.05$} & {\scriptsize 1 + 0.571429 max(0, 4 + n - x)} & {\scriptsize $6$} & {\scriptsize $0.27$} & {\scriptsize ((x * -4/7) + (n * 4/7) + 37/21)}\\ 
	\midrule  
	prseq & {\scriptsize $0.04$} & {\scriptsize 0.65 max(0, x - y) + 0.35 max(0, y)} & {\scriptsize $266$} & {\scriptsize $13.63$} & {\scriptsize ((x * 13/20) + (y * -1/2) + 47/20)}\\ 
	\midrule  
	prspeed & {\scriptsize $0.04$} & {\scriptsize  \makecell[cc]{1 + 2 max(0, m - y) \\+ 0.666667 max(0, n - x)}} & {\scriptsize $18$} & {\scriptsize $0.41$} & {\scriptsize \makecell[cc]{max[ ((y * -2.0) + (m * 2.0) + (n * 2/3) + 1/3),\\ ((x * -2/3) + (n * 2/3) + 1/3) ]}}\\ 
	\midrule  
	race & {\scriptsize $0.17$} & {\scriptsize 1 + 0.666667 max(0, 9 - h + t)} & {\scriptsize $32$} & {\scriptsize $1.95$} & {\scriptsize ((h * -2/3) + (t * 2/3) + 13/3)}\\ 
	\midrule  
	rejectionsampling & {\scriptsize $4.03$} & {\scriptsize 1 + 5 max(0, n)} & {\scriptsize $20$} & {\scriptsize $0.53$} & {\scriptsize ((n * 5.0) + 1.0)}\\ 
	\midrule  
	rfindmc & {\scriptsize ${\leq}0.01$} & {\scriptsize 1 + max(0, -i + k)} & {\scriptsize $1$} & {\scriptsize $0.07$} & {\scriptsize (k * 2.0)}\\ 
	\midrule  
	rfindlv & {\scriptsize ${\leq}0.01$} & {\scriptsize 1 + 2 max(0, flag)} & {\scriptsize $1$} & {\scriptsize $0.05$} & {\scriptsize 3.0}\\ 
	\midrule  
	rdseql & {\scriptsize $0.02$} & {\scriptsize 1 + 2.25 max(0, x) + max(0, y)} & {\scriptsize $13$} & {\scriptsize $0.29$} & {\scriptsize ((x * 9/4) + y + -1/4)}\\ 
	\midrule  
	rdspeed & {\scriptsize $0.03$} & {\scriptsize \makecell[cc]{1 + 2 max(0, m - y) \\+ 0.666667 max(0, n - x)}} & {\scriptsize $18$} & {\scriptsize $0.44$} & {\scriptsize \makecell[cc]{ max[ ((y * -2.0) + (m * 2.0) + (n * 2/3) + 1/3),\\ ((x * -2/3) + (n * 2/3) + 1/3) ]}}\\ 
	\midrule  
	sprdwalk & {\scriptsize ${\leq}0.01$} & {\scriptsize 1 + 2 max(0, n - x)} & {\scriptsize $3$} & {\scriptsize $0.08$} & {\scriptsize ((x * -3.0) + (n * 3.0))}\\ 
	\midrule  
\end{tabular}
	}
\end{table}
\FloatBarrier

\subsection{Details on the Comparison of \toolname{Exist} and \tool}
\label{sec:exist}

\Cref{tab:exist} depicts the results on the sub-invariant benchmarks. \headername{Prog} is the name of the benchmark, t denotes the runtime in seconds required by the tools, $I$ denotes the sub-invariants computed by each tool, and $|S'|$ depicts the number of counterexamples required by \tool.

\begin{table}[t]
	\centering
	\caption{\toolname{Exist} vs.\ \tool (TO=5min, MO=8GB. Time in seconds).}\label{tab:exist}
	\adjustbox{width=1.0\textwidth}{

\begin{tabular}{l||rc||rrc}
	& \multicolumn{2}{c}{\headername{Exist}} &  \multicolumn{3}{c}{\tool} \\ 
	Prog & {\footnotesize t} & $I$ & {\footnotesize $|S'|$} & {\footnotesize t} & $I$ \\ 
	\midrule\midrule 
	BiasDir1\_0 &{\scriptsize $78.31$} &{\scriptsize  $x + [x = y] \cdot   (-0.5\cdot x + -0.5\cdot y + 0.1)$ }  &{\scriptsize $0$} & {\scriptsize $0.15$} & {\scriptsize $\iverson{\text{loop guard}}\cdot 0 + \iverson{\neg \text{loop guard}}\cdot x$}\\ 
	BiasDir1\_1 &{\scriptsize $97.29$}&{\scriptsize  $x + [x = y] \cdot   (-0.5\cdot x + -0.5\cdot y + 0.5)$ }& {\scriptsize $1$} & {\scriptsize $0.08$} & {\scriptsize $\iverson{\text{loop guard}}\cdot 0.5 + \iverson{\neg \text{loop guard}}\cdot x$} \\ 
	BiasDir2\_0 &{\scriptsize $66.42$} &{\scriptsize  $x + [x = y] \cdot   (-0.5\cdot x + -0.5\cdot y + 0.1)$ } & {\scriptsize $0$} & {\scriptsize $0.07$} & {\scriptsize $\iverson{\text{loop guard}}\cdot 0 + \iverson{\neg \text{loop guard}}\cdot x$}\\ 
	BiasDir2\_1 &{\scriptsize $171.83$} & {\scriptsize $x + [x = y] \cdot   (-0.5\cdot x + -0.5\cdot y + 0.5)$ }& {\scriptsize $1$} & {\scriptsize $0.08$} & {\scriptsize $\iverson{\text{loop guard}}\cdot 0.5 + \iverson{\neg \text{loop guard}}\cdot x$}\\ 
	BiasDir3\_0 &{\scriptsize $66.55$} &{\scriptsize  $x + [x = y] \cdot   (-0.3\cdot x + -0.3\cdot y)$ }&{\scriptsize $0$} & {\scriptsize $0.08$} & {\scriptsize $\iverson{\text{loop guard}}\cdot 0 + \iverson{\neg \text{loop guard}}\cdot x$}\\ 
	BiasDir3\_1 &{\scriptsize $99.23$} & {\scriptsize $x + [x = y] \cdot   (-0.5\cdot x + -0.5\cdot y + 0.5)$ }& {\scriptsize $1$} & {\scriptsize $0.08$} & {\scriptsize $\iverson{\text{loop guard}}\cdot 0.5 + \iverson{\neg \text{loop guard}}\cdot x$}\\ 
	Bin01\_0 &{\scriptsize $53.19$} &{\scriptsize  $x + [n > 0] \cdot  0$ }& {\scriptsize $1$} & {\scriptsize $0.06$} & {\scriptsize $\iverson{\text{loop guard}}\cdot x + \iverson{\neg \text{loop guard}}\cdot x$}\\ 
	Bin02\_0 &{\scriptsize $83.04$} &{\scriptsize  $x + [n > 0] \cdot  0$ }& {\scriptsize $1$} & {\scriptsize $0.06$} & {\scriptsize $\iverson{\text{loop guard}}\cdot x + \iverson{\neg \text{loop guard}}\cdot x$}\\ 
	Bin03\_0 &{\scriptsize $53.13$} &{\scriptsize  $x + [n > 0] \cdot  0$ }& {\scriptsize $1$} & {\scriptsize $0.06$} & {\scriptsize $\iverson{\text{loop guard}}\cdot x + \iverson{\neg \text{loop guard}}\cdot x$}\\ 
	Bin11\_0 &{\scriptsize $32.52$} &{\scriptsize   $x + [n < M] \cdot  0$ }& {\scriptsize $1$} & {\scriptsize $0.06$} & {\scriptsize $\iverson{\text{loop guard}}\cdot x + \iverson{\neg \text{loop guard}}\cdot x$} \\ 
	Bin11\_1 &{\scriptsize $\textnormal{TO}$} &{\scriptsize  -- }& {\scriptsize $3$} & {\scriptsize $0.08$} & {\scriptsize $\iverson{\text{loop guard}}\cdot (-0.5*n +x +1/2*M) + \iverson{\neg \text{loop guard}}\cdot x$}\\ 
	Bin12\_0 &{\scriptsize $78.19$} & {\scriptsize $x + [n < M] \cdot  0$ }& {\scriptsize $1$} & {\scriptsize $0.06$} & {\scriptsize $\iverson{\text{loop guard}}\cdot x + \iverson{\neg \text{loop guard}}\cdot x$}\\ 
	Bin12\_1 &{\scriptsize $218.17$} &{\scriptsize  $x + [n < M] \cdot   (-0.1\cdot n + 0.1\cdot M)$}& {\scriptsize $3$} & {\scriptsize $0.09$} & {\scriptsize $\iverson{\text{loop guard}}\cdot (-0.1*n +x +0.1*M) + \iverson{\neg \text{loop guard}}\cdot x$}\\ 
	Bin13\_0 &{\scriptsize $32.95$} & {\scriptsize $x + [n < M] \cdot  0$}& {\scriptsize $1$} & {\scriptsize $0.05$} & {\scriptsize $\iverson{\text{loop guard}}\cdot x + \iverson{\neg \text{loop guard}}\cdot x$}\\ 
	Bin13\_1 &{\scriptsize $\textnormal{TO}$} &{\scriptsize  -- }&  {\scriptsize $3$} & {\scriptsize $0.08$} & {\scriptsize $\iverson{\text{loop guard}}\cdot (-0.9*n+x+0.9*M) + \iverson{\neg \text{loop guard}}\cdot x$}\\ 
	Bin21\_0 &{\scriptsize $54.12$} &{\scriptsize  $x + [n > 0] \cdot  0$ }& {\scriptsize $2$} & {\scriptsize $0.06$} & {\scriptsize $\iverson{\text{loop guard}}\cdot x + \iverson{\neg \text{loop guard}}\cdot x$}\\ 
	Detm1\_0 &{\scriptsize $22.8$} &{\scriptsize  $count + [x <= 10] \cdot  0$ }& {\scriptsize $3$} & {\scriptsize $0.06$} & {\scriptsize $\iverson{\text{loop guard}}\cdot count + \iverson{\neg \text{loop guard}}\cdot count$}\\ 
	Detm1\_1 &{\scriptsize $57.72$} & {\scriptsize $count + [x <= 10] \cdot   (1)$ }& {\scriptsize $3$} & {\scriptsize $0.08$} & {\scriptsize $\iverson{\text{loop guard}}\cdot (count + 1) + \iverson{\neg \text{loop guard}}\cdot count$}\\ 
	Duel1\_0 &{\scriptsize $\textnormal{TO}$} &{\scriptsize  -- }& {\scriptsize $5$} & {\scriptsize $0.12$} & {\scriptsize $\iverson{\text{loop guard}}\cdot \nicefrac{10}{19} + \iverson{\neg \text{loop guard}}\cdot t$}\\ 
	Duel2\_0 &{\scriptsize $\textnormal{TO}$} &{\scriptsize  --} &  {\scriptsize $1$} & {\scriptsize $0.09$}& {\scriptsize $\iverson{\text{loop guard}}\cdot \nicefrac{10}{11} + \iverson{\neg \text{loop guard}}\cdot t$}\\ 
	Fair1\_0 &{\scriptsize $30.29$} & {\scriptsize $count + [ c1 + c2 = 0] \cdot  0$ }& {\scriptsize $2$} & {\scriptsize $0.08$} & {\scriptsize $\iverson{\text{loop guard}}\cdot count + \iverson{\neg \text{loop guard}}\cdot count$}\\ 
	Fair1\_1 &{\scriptsize $37.05$} &{\scriptsize  $count + [ c1 + c2 = 0] \cdot   (1.0\cdot 1)$} & {\scriptsize $2$} & {\scriptsize $0.1$} & {\scriptsize $\iverson{\text{loop guard}}\cdot (count +1) + \iverson{\neg \text{loop guard}}\cdot count$}\\ 
	Gambler01\_0 &{\scriptsize $35.81$} & {\scriptsize $z + [(x > 0) \& (y > x)] \cdot  0$ }& {\scriptsize $2$} & {\scriptsize $0.07$} & {\scriptsize $\iverson{\text{loop guard}}\cdot z + \iverson{\neg \text{loop guard}}\cdot z$}\\ 
	Geo01\_0 &{\scriptsize $31.71$} &{\scriptsize  $z + [flip = 0] \cdot  0$} & {\scriptsize $1$} & {\scriptsize $0.05$} & {\scriptsize $\iverson{\text{loop guard}}\cdot z + \iverson{\neg \text{loop guard}}\cdot z$}\\ 
	Geo01\_1 &{\scriptsize $68.53$} & {\scriptsize $z + [flip = 0] \cdot   (0.5 \cdot 1)$ }& {\scriptsize $1$} & {\scriptsize $0.05$} & {\scriptsize $\iverson{\text{loop guard}}\cdot (0.5*z+0.5) + \iverson{\neg \text{loop guard}}\cdot z$}\\ 
	Geo01\_2 &{\scriptsize $73.85$} & {\scriptsize $z + [flip = 0] \cdot   (1.0 \cdot 1)$} & {\scriptsize $2$} & {\scriptsize $0.09$} & {\scriptsize $\iverson{\text{loop guard}}\cdot (z+0.5) + \iverson{\neg \text{loop guard}}\cdot z$}\\ 
	Geo11\_0 &{\scriptsize $42.13$} &{\scriptsize  $z + [flip = 0] \cdot  0$ } & {\scriptsize $1$} & {\scriptsize $0.05$} & {\scriptsize $\iverson{\text{loop guard}}\cdot z + \iverson{\neg \text{loop guard}}\cdot z$}\\ 
	Geo21\_0 &{\scriptsize $37.1$} &{\scriptsize  $z + [flip = 0] \cdot  0$ } &{\scriptsize $1$} & {\scriptsize $0.05$} & {\scriptsize $\iverson{\text{loop guard}}\cdot z + \iverson{\neg \text{loop guard}}\cdot z$}\\ 
	GeoAr01\_0 &{\scriptsize $122.83$} & {\scriptsize $x + [z != 0] \cdot   (1.0\cdot y)$ } &{\scriptsize $2$} & {\scriptsize $0.06$} & {\scriptsize $\iverson{\text{loop guard}}\cdot x + \iverson{\neg \text{loop guard}}\cdot x$}\\ 
	GeoAr01\_1 &{\scriptsize $33.81$} &{\scriptsize  $x + [z != 0] \cdot  0$ } & {\scriptsize $2$} & {\scriptsize $0.08$} & {\scriptsize $\iverson{\text{loop guard}}\cdot (x+y) + \iverson{\neg \text{loop guard}}\cdot z$}\\ 
	LinExp1\_0 &{\scriptsize $189.11$} &{\scriptsize  $z + [n > 0] \cdot   (n^{1.0}\cdot 2)$ }&{\scriptsize $7$} & {\scriptsize $0.53$} & {\scriptsize $\iverson{\text{loop guard}}\cdot (\nicefrac{16}{21}*z + 2*n) + \iverson{\neg \text{loop guard}}\cdot z$}\\  
	LinExp1\_1 &{\scriptsize $196.32$} &{\scriptsize  $z + [n > 0] \cdot   (1.0\cdot n + 1.0\cdot 1)$ }&{\scriptsize $3$} & {\scriptsize $0.52$} & {\scriptsize $\iverson{\text{loop guard}}\cdot (z+2) + \iverson{\neg \text{loop guard}}\cdot z$}\\  
	PrinSys1\_0 &{\scriptsize $13.27$} &{\scriptsize  $[x=1] \cdot  1 + [x = 0] \cdot  0$ } &{\scriptsize $0$} & {\scriptsize $0.06$} & {\scriptsize $[x=1]*1 + [x \neq 1]*0$}\\ 
	RevBin1\_0 &{\scriptsize $179.97$} &{\scriptsize $z + [x > 0] \cdot   (x^{1}\cdot 2)$ } &{\scriptsize $2$} & {\scriptsize $0.09$}& \scriptsize{$\iverson{\text{loop guard}}\cdot (x+z) + \iverson{\neg \text{loop guard}}\cdot z$}\\ 
	RevBin1\_1 &{\scriptsize $52.03$} & {\scriptsize $z + [x > 0] \cdot  0$ } & {\scriptsize $1$} & {\scriptsize $0.05$}& \scriptsize{$\iverson{\text{loop guard}}\cdot z + \iverson{\neg \text{loop guard}}\cdot z$}\\ 
	Sum01\_0 &{\scriptsize $\textnormal{TO}$} &{\scriptsize  -- } &{\scriptsize $4$} & {\scriptsize $0.09$} & \scriptsize{$\iverson{\text{loop guard}}\cdot (0.25*n +x) + \iverson{\neg \text{loop guard}}\cdot x$} \\ 
	Mart1\_0 &{\scriptsize $84.39$} & {\scriptsize $rounds + [b > 0] \cdot  0$ } & {\scriptsize $-$} & {\scriptsize $\textnormal{TO}$} & -- \\ 
	Mart1\_1 &{\scriptsize $\textnormal{TO}$} &{\scriptsize  -- }&{\scriptsize $-$} & {\scriptsize $\textnormal{TO}$}& -- \\ 
\end{tabular}
}
\end{table}

\FloatBarrier

\subsection{Programs from \Cref{tab:storm}}
\label{sec:programs_storm}

This secton contains the programs from \Cref{tab:storm}.
%
%
\begin{figure}[h]
	
 \begin{lstlisting}
nat x [0,20000];
nat s [1,5];

while(0<x & x<20000 & 1<=s & s <=  5){
   {x:=x-1}
   [0.5]
   {
   if (x=1){
      s := 1 : 1/5 + 2 : 1/5 + 3 : 1/5 + 4: 1/5 + 5 : 1/5;
   }else{
      skip
   };
    x := x + s;
  }
}
\end{lstlisting}
\caption{boundedrwmultistep}
\end{figure}

\begin{figure}[h]
	
	\begin{lstlisting}
	nat sent [0,8000000000];
	nat failed [0,10];
	
	while(failed<10 & sent<8000000000){
	   {failed:=0;sent:=sent+1}[0.99]{failed:=failed+1}
	}
	\end{lstlisting}
	\caption{brp}
\end{figure}

\begin{figure}[h]
	
	\begin{lstlisting}
	nat sent [0,8000000];
	nat failed [0,5];
	nat MAXSENT [0,8000000];
	nat MINFAILED;
	
	while(failed<MINFAILED & sent<MAXSENT & MAXSENT <= 8000000 & 5 <= MINFAILED){
	   {failed:=0;sent:=sent+1}[0.99]{failed:=failed+1}
	}
	\end{lstlisting}
	\caption{brpfinitefamily}
\end{figure}

\begin{figure}[h]
	
	\begin{lstlisting}
	nat c [0,1];
	nat x [0,1000000000000];
	
	while(c<=0 & x<1000000000000){
	   {c:=1}[0.000000000001]{x:=x+1}
	}
	
	\end{lstlisting}
	\caption{chain}
\end{figure}

\begin{figure}[h]
	
	\begin{lstlisting}
	nat c [0,1];
	nat x [0,10000000];
	nat step [0,10];
	
	while(c<=0 & x<10000000 & 0<=step & step <=10){
	   if(step=0){
	      step:=  (1) : 1/10 + (2) : 1/10 + (3) : 1/10 + (4) : 1/10 
	                    + (5) : 1/10+ (6) : 1/10+ (7) : 1/10+ (8) : 1/10 
	                    + (9) : 1/10 + (10) : 1/10;
	   }else{
	      if(step <= 2){
	         {c:=1}[0.0000001]{x:=x+step}
	      }else{
	         if(step <= 4){
	            {c:=1}[0.0000002]{x:=x+step}
	         }else{
	            if(step <=6){
	               {c:=1}[0.0000003]{x:=x+step}
	            }else{
	               if(step <=8){
	                  {c:=1}[0.0000004]{x:=x+step}
	               }else{
	                  {c:=1}[0.0000005]{x:=x+step}
	               }
	            }
	         }
	      }
	    }
	}
	\end{lstlisting}
	\caption{chainselectstepsize}
\end{figure}

\begin{figure}[h]
	\begin{lstlisting}
	nat a [0,1000];
	nat b [0,1000];
	
	while(a<1000 & b<1000){
	   {a:=a+1}[0.5]{b:=b+1}
	}
	\end{lstlisting}
	\caption{gridbig}
\end{figure}

\begin{figure}[h]
	\begin{lstlisting}
	nat a [0,10];
	nat b [0,10];
	
	while(a<10 & b<10){
	   {a:=a+1}[0.5]{b:=b+1}
	}
	\end{lstlisting}
	\caption{gridsmall}
\end{figure}

\begin{figure}[h]
	\begin{lstlisting}
	nat start [0,1];
	nat established [0,1];
	nat curprobe [0,100000000];
	
	while(curprobe < 100000000 & established <=0 & start <= 1){
	   if(start = 1){
	      {start:=0} [0.5] {start:=0; established:=1}
	   }else{ 
	      {curprobe := curprobe + 1} 
	          [0.999999999]  {start:=1;curprobe:=0} 
	   }
	}
	\end{lstlisting}
	\caption{zeroconf}
\end{figure}

\begin{figure}[h]
	\begin{lstlisting}
	nat start [0,1];
	nat established [0,1];
	nat curprobe [0,200000000];
	nat N [100000000,200000000];
	
	while(curprobe < N & established <=0 & start <= 1 & 100000000 <= N & N <= 200000000){
	
	   if(start = 1){
	      {start:=0} [0.5] {start:=0; established:=1}   
	   }else{ 
	      {curprobe := curprobe + 1} 
	         [0.999999999]  {start:=1;curprobe:=0} }
	   }
	\end{lstlisting}
	\caption{zeroconffamily}
\end{figure}

\end{document}